\documentclass{aastex62}
\usepackage{graphicx}
\usepackage[caption=false]{subfig}
\pdfoutput=0

\def\mnras{{MNRAS}}
\def\apj{{ApJ}}
\def\aj{{AJ}}
\def\aap{{A\&Ap}}
\def\apjl{{ApJL}}
\def\apjs{{ApJS}}
\def\araa{{ARAA}}
\def\nat{{Nature}}

\def\Chandra{{\em Chandra}}

\newcommand{\aox}{\ifmmode{\alpha_{\mathrm{ox}}} \else $\alpha_{\mathrm{ox}}$\fi} 
\newcommand{\atoms}{\ifmmode{\mathrm{\,atoms~cm^{-2}}} \else \,atoms cm$^{-2}$\fi}
\newcommand{\ax}{\ifmmode{\alpha_x} \else $\alpha_x$\fi} 
\newcommand{\cmsq}{\ifmmode{\mathrm{cm^{-2}}} \else cm$^{-2}$\fi}
\newcommand{\degs}{\ifmmode ^{\circ}\else$^{\circ}$\fi}
\newcommand{\degsq}{\ifmmode {\mathrm{deg^2}} \else deg$^2$\fi}
\newcommand{\perdegsq}{\ifmmode {\mathrm{deg^{-2}}} \else deg$^{-2}$\fi}
\newcommand{\ew}{\ifmmode{W_{\lambda}} \else $W_{\lambda}$\fi}
\newcommand{\fbol}{\ifmmode f_{\mathrm{bol}} \else $f_{\mathrm{bol}}$\fi} 
\newcommand{\fcgs}{\ifmmode \mathrm{erg~cm^{-2}~s^{-1}}\else erg~cm$^{-2}$~s$^{-1}$\fi}
\newcommand{\flamcgs}{\ifmmode \mathrm{erg\,cm^{-2}\,s^{-1}\,\AA^{-1}}\else erg\,cm$^{-2}$\,s$^{-1}$\,\AA$^{-1}$)\fi}
\newcommand{\fnucgs}{\ifmmode {\mathrm{erg~cm^{-2}~s^{-1}~Hz^{-1}}}\else erg~cm$^{-2}$~s$^{-1}$~Hz$^{-1}$\fi}
\newcommand{\gax }{{\lower0.8ex\hbox{$\buildrel >\over\sim$}}}
\newcommand\Ha{\ifmmode {\mathrm H}\alpha \else H$\alpha$\fi}
\newcommand\Hb{\ifmmode {\mathrm H}\beta \else H$\beta$\fi}
\newcommand{\kms}{\ifmmode~{\mathrm{km~s}}^{-1}\else ~km~s$^{-1}~$\fi}
\newcommand{\lax }{{\lower0.8ex\hbox{$\buildrel <\over\sim$}}}
\newcommand{\lcgs}{\ifmmode \mathrm{erg~s^{-1}}\else erg~s$^{-1}$\fi}
\newcommand{\lnucgs}{\ifmmode erg~s^{-1}~Hz^{-1}\else erg~s$^{-1}$~Hz$^{-1}$\fi}

\newcommand{\logz}{\ifmmode{\mathrm{log}}~z \else log$~z$\fi}
\newcommand{\lo}{\ifmmode l_o \else $~l_o$\fi}
\newcommand{\Lo}{\ifmmode L_o \else $~L_o$\fi}
\newcommand{\lx}{\ifmmode l_x \else $~l_x$\fi}
\newcommand{\Lx}{\ifmmode L_x \else $~L_x$\fi}
\newcommand{\lbol}{\ifmmode L_{\mathrm{bol}} \else $L_{\mathrm{bol}}$\fi}
\newcommand{\Lbol}{\ifmmode L_{\mathrm{bol}} \else $L_{\mathrm{bol}}$\fi}
\newcommand{\LBol}{\ifmmode L_{\mathrm{bol}} \else $L_{\mathrm{bol}}$\fi}
\newcommand{\LEdd}{\ifmmode L_{\mathrm{Edd}} \else $L_{\mathrm{Edd}}$\fi}
\newcommand{\LxLbol}{\ifmmode L_x/L_{\mathrm{bol}} \else $L_x/L_{\mathrm{bol}}$\fi}
\newcommand{\rEdd}{\ifmmode L/L_{\mathrm{Edd}} \else $L/L_{\mathrm{Edd}}$\fi}
\newcommand{\REdd}{\ifmmode L/L_{\mathrm{Edd}} \else $L/L_{\mathrm{Edd}}$\fi}
\newcommand{\Rblr}{\ifmmode {R_{\mathrm BLR}} \else $R_{\mathrm BLR}$\fi}
\newcommand{\lamEdd}{\ifmmode \lambda_{\mathrm{Edd}} \else $\lambda_{\mathrm{Edd}}$\fi}
\newcommand{\mbh}{\ifmmode {M_{\rm BH}}\else${M_{\rm BH}}$\fi}
\newcommand{\Mbh}{\ifmmode {M_{\rm BH}}\else${M_{\rm BH}}$\fi}
\newcommand{\mdot}{\ifmmode \dot{m} \else $\dot{m}$\fi}
\newcommand{\mdote}{\ifmmode \dot{m}_{E} \else $\dot{m}_{E}$\fi}
\newcommand{\mone}{\ifmmode ^{-1}\else$^{-1}$\fi}
\newcommand{\msun}{\ifmmode {M_{\odot}}\else${M_{\odot}}$\fi}
\newcommand{\Msun}{\ifmmode {M_{\odot}}\else${M_{\odot}}$\fi}
\newcommand{\mtwo}{\ifmmode ^{-2}\else$^{-2}$\fi}
\newcommand{\Mvir}{\ifmmode {M_{\rm BH}^{\mathrm SE}}\else${M_{\rm BH}^{\mathrm SE}}$\fi}
\newcommand{\nhgal}{\ifmmode{ N_{H}^{Gal}} \else N$_{H}^{Gal}$\fi}
\newcommand{\nh}{\ifmmode{\mathrm N_{H}} \else N$_{H}$\fi}
\newcommand{\nhintr}{\ifmmode{ N_{H}^{intr}} \else N$_{H}^{intr}$\fi}
\newcommand{\nhtot}{\ifmmode{ N_{H}^{tot}} \else N$_{H}^{tot}$\fi}
\newcommand{\nhz}{\ifmmode{ N_{H}^z} \else N$_{H}^z$\fi}
\newcommand{\oi}{\ifmmode{\mathrm [O\,II]} \else [O\,II]\fi}
\newcommand{\oii}{\ifmmode{\mathrm [O\,II]} \else [O\,II]\fi}
\newcommand{\oiii}{\ifmmode{\mathrm [O\,III]} \else [O\,III]\fi}
\newcommand{\optebl}{\ifmmode L_{\rm 2500\,\AA} \else $~L_{\rm 2500\,\AA}$\fi}
\newcommand{\opteml}{\ifmmode l_{\mathrm{2500\,\AA}} \else $~l_{\mathrm{2500\,\AA}}$\fi}
\newcommand{\Teff}{\ifmmode T_{\mathrm{Eff}} \else $T_{\mathrm{Eff}}$\fi}
\newcommand{\xebl}{\ifmmode L_X \else $~L_X$\fi}
\newcommand{\xeml}{\ifmmode l_{\mathrm{2\,keV}} \else $~l_{\mathrm{2\,keV}}$\fi}
\newcommand{\XMM}{XMM-{\em Newton}}
\def\geqsim{\lower.73ex\hbox{$\sim$}\llap{\raise.4ex\hbox{$>$}}$\,$}
\def\leqsim{\lower.73ex\hbox{$\sim$}\llap{\raise.4ex\hbox{$<$}}$\,$}

\newcommand{\umg}{\ifmmode{\mathrm{(}u-g\mathrm{)}} \else ($u-g$)\fi}
\newcommand{\gmr}{\ifmmode{\mathrm{(}g-r\mathrm{)}} \else ($g-r$)\fi}
\newcommand{\rmi}{\ifmmode{\mathrm{(}r-i\mathrm{)}} \else ($r-i$)\fi}
\newcommand{\gmi}{\ifmmode{\mathrm{(}g-i\mathrm{)}} \else ($g-i$)\fi}
\newcommand{\imz}{\ifmmode{\mathrm{(}i-z\mathrm{)}} \else ($i-z$)\fi}
\newcommand{\jmh}{\ifmmode{\mathrm{(}J-H\mathrm{)}} \else ($J-H$)\fi}
\newcommand{\hmk}{\ifmmode{\mathrm{(}H-K\mathrm{)}} \else ($H-K$)\fi}
\newcommand{\ctwo}{\ifmmode C_2 \else C$_2$\fi}

\graphicspath{{./}}

\received{May 15, 2019}
\revised{June 20, 2019}
\shorttitle{Dwarf Carbon Stars - Rejuvenated by Accretion?}
\shortauthors{Green et al.}
\begin{document}

\title{A Chandra Study: Are Dwarf Carbon Stars Spun Up and Rejuvenated by Mass Transfer?}

\correspondingauthor{Paul J. Green}
\email{pgreen@cfa.harvard.edu}

\author[0000-0002-8179-9445]{Paul J. Green}
\affil{Harvard Smithsonian Center for Astrophysics, 60 Garden St, Cambridge, MA 02138, USA}
\author[0000-0002-6752-2909]{Rodolfo Montez}
\affil{Harvard Smithsonian Center for Astrophysics, 60 Garden St, Cambridge, MA 02138, USA}
%\collaboration{(AAS Journals Data Scientists collaboration)}
\author{Fernando Mazzoni}
\affiliation{University of Massachusetts, Lowell}
\author[0000-0002-0201-8306]{Joseph Filippazzo}
\affiliation{Space Telescope Science Institute 3700 San Martin Drive, Baltimore, MD 21218, USA}
\author[0000-0002-6404-9562]{Scott F. Anderson}
\affiliation{Department of Astronomy, University of Washington, Box 351580, Seattle, WA 98195, USA}
\author[0000-0002-1126-869X]{Orsola De Marco}
\affiliation{Department of Physics \& Astronomy, Macquarie University, Sydney, NSW 2109, Australia; Astronomy, Astrophysics and Astrophotonics Research Centre, Macquarie University, Sydney, NSW 2109, Australia}
\author[0000-0002-0210-2276]{Jeremy J. Drake}
\affil{Harvard Smithsonian Center for Astrophysics, 60 Garden St, Cambridge, MA 02138, USA}
\author[0000-0003-1748-602X]{Jay Farihi}
\affiliation{Department of Physics and Astronomy, University College London, WC1E 6BT, UK}
\author{Adam Frank}
\affiliation{Department of Physics and Astronomy, University of Rochester, Rochester, NY 14627-0171, USA}
\author[0000-0002-3138-8250]{Joel H. Kastner}
\affiliation{Chester F. Carlson Center for Imaging Science, Rochester Institute of Technology, 54 Lomb Memorial Drive, Rochester NY 14623 USA ; School of Physics \& Astronomy and Laboratory for Multiwavelength Astrophysics, Rochester Institute of Technology, 54 Lomb Memorial Drive, Rochester NY 14623 USA)}
\author[0000-0003-2561-6306]{Brent Miszalski}
\affiliation{South African Astronomical Observatory, PO Box 9, Observatory 7935, South Africa; Southern African Large Telescope Foundation, PO Box 9, Observatory 7935, South Africa}
\author[0000-0002-9453-7735]{Benjamin R. Roulston}
\affiliation{Department of Astronomy, Boston University, 725 Commonwealth Avenue, Boston, MA 02215, USA}
\affil{Harvard Smithsonian Center for Astrophysics, 60 Garden St, Cambridge, MA 02138, USA}

%\nocollaboration



\begin{abstract}
 Carbon stars (with C/O$>1$) were long assumed to all be giants, because
only AGB stars dredge up significant carbon into their atmospheres.  The case is nearly iron-clad now that the formerly mysterious dwarf carbon (dC) stars are actually far more common than C giants, and have accreted carbon-rich material from a former AGB companion, yielding a white dwarf and a dC star that has gained both significant mass and angular momentum. Some such dC systems have undergone a planetary nebula phase, and some may evolve to become CH, CEMP, or Ba giants. Recent studies indicate that most dCs are likely from older, metal-poor kinematic populations.  Given the well-known anti-correlation of age and activity, dCs would not be expected to show significant X-ray emission related to coronal activity. However,  accretion spin-up might be expected to rejuvenate magnetic dynamos in these post mass-transfer binary systems.  We describe our \Chandra\, pilot study of six dCs selected from the SDSS for H$\alpha$ emission and/or a hot white dwarf companion, to test whether their X-ray emission strength and spectral properties are consistent with a rejuvenated dynamo.  We detect all 6 dCs in the sample, which have X-ray luminosities ranging from log$\Lx \sim 28.5$ -- $29.7$, preliminary evidence that dCs may be active at a level consistent with stars that have short rotation periods of several days or less.  More definitive results require a sample of typical dCs with deeper X-ray observations to better constrain their plasma temperatures.
\end{abstract}


%% We recommend that authors also use the nat \citep

\section{Introduction} 
\label{sec:intro}

\subsection{Dwarf Carbon Stars}

Carbon (C) stars show molecular absorption bands of carbon --- C$_2$, CN
or CH in their optical spectra --- because they have C/O$>1$, with atmospheres cool enough to form molecules.   
By contrast, the C/O ratio for nearby Sun-like stars is between $\sim$0.5 -- 0.8 (\citealt{Fortney12} and references therein). In most single stars of intermediate mass  ($\sim 1 - 8\,\msun$), 
atmospheres show C/O above unity only in the asymptotic giant branch phase, when carbon is brought 
to the surface during episodes of strong convection.   The so-called third
dredge-up occurs during late stages of thermal pulsations (the TP-AGB phase), associated with the triple-$\alpha$ process in shell He-burning episodes. AGB stars may reach a radius of $\sim 1\,$A.U. ($\sim$200\,R$_{\odot}$).  Pulsations, shocks and radiation all drive carbon- and $s$-process- enhanced material outward in an AGB wind, which may result in detached shells of circumstellar material containing 50 -- 70\% of the star's mass \citep{Wood04}.  Winds from AGB stars are the predominant source of carbon in the interstellar medium, the PAHs in star-forming regions, pre-biotic molecules and indeed the planets as well (e.g., \citealt{Tielens05}).  

% I HAD \bf FOR ALL NEW TEXT IN RESPONSE TO REFEREE, WHICH I KEPT FOR A VERSINO TO TRACK CHANGES.  I JUST MODIFIED THAT TO BE \, FOR THE RESUBMITTED VERSION.

The discovery of the first {\em main sequence} carbon star G77-61
($M_V=10.1$; \citealt{Dahn77}) sounded like an oxymoron at the time.  {\, \cite{Dahn77} offered several possible explanations for the existence of a dwarf carbon (dC) star.}
That G77-61 is in a binary system ($P\sim 245$d; \citealt{Dearborn86}) with an unseen companion consistent with a cool white dwarf, strongly suggested that extrinsic processes produce the dC's enhanced atmospheric
abundances.  {\, Recent results from radial velocity monitoring of larger samples of dCs are consistent with a 100\% binary fraction \citep{Whitehouse18, Roulston19}.  Therefore,} the favored hypothesis {\, to explain dCs} is that C-rich material lost in an
AGB wind can be efficiently captured by a main-sequence companion. While the AGB donor has since evolved to a white dwarf (WD), the
``polluted'' companion, as an innocent bystander, could be either
dwarf or giant in the current epoch.  {\, Indeed,} a variety of non-AGB stars, including many red giant stars, show enhanced  carbon and/or \hbox{$s$-process} abundances that are similarly extrinsic.  The CH, Ba and the
carbon-enhanced metal poor (CEMP-$s$) stars \citep{Lucatello05} {\, - mostly giants or subgiants - } likely evolved from dC stars, and have been more commonly studied  only by virtue of their greater brightness.  {\, All these objects can show atmospheric signatures of \hbox{$s$-process} elements associated with neutron capture onto iron seed nuclei in AGB stars.

\,{ Is there another viable path to strong carbon enhancement besides mass transfer? Intriguingly, the CEMP-no class lacks \hbox{$s$-process} overabundances, and initial radial velocity studies do not show an enhanced binary fraction \citep{Starkenburg14}, so CEMP-no stars may achieve atmospheric carbon enhancement without mass transfer.  Debate continues, but one suggestion is that (reminiscent of one of the suggestions in \citealt{Dahn77}) CEMP-no stars may have coalesced from clouds seeded with strong carbon overabundances (and insignificant iron) by the first zero-heavy-element stars (e.g., \citealt{Bromm03, Norris13}).  This interpretation is favored by the fact that nearly all CEMP-no stars have [Fe/H]$\leq -4.0$.}  {\, The prototype dC star}, G77-61 has a companion, but also has extremely low metallicity ([Fe/H]$\sim -4$; \citealt{Plez05}), so might be a candidate for either mechanism.} 

The importance of dC stars like G77-61 was later confirmed as many additional dCs
were recognized from their large proper motions \citep{Green91,Downes04}. In a handful of cases, the ``smoking gun'' of AGB binary mass transfer was revealed as a hot DA 
white dwarf companion \citep{Heber93,Liebert94}.  More recently \citet{Green13} found more than 700 dC stars in the SDSS,
powerfully confirming that dCs are in fact the {\em numerically dominant} type of carbon star in the Galaxy, supporting the ubiquity
of the dC phenomenon across a broad range of mass and age. Other large samples have since emerged from LAMOST \citep{Ji16,Li18}, including
a dozen double-line spectroscopic binary (SB2) DA+dC spectra, which
reveal recently-minted systems \Citep{Si14}.  
% While the fraction of AGB systems that create a dC is not known, the large space density of dCs compared to C-AGB stars is surely due to the extreme brevity of the AGB phase relative to main sequence lifetimes.
Dwarf carbon stars show kinematic evidence of mostly belonging to older populations \citep{Green13,Farihi18}.  Dwarf carbon stars may be favored in older e.g., thick disk or halo systems {\, because they are born with large C/O (as for the CEMP-no hypothesis above) or simply} because at lower metallicity, less mass transfer is required to boost C/O$>1$.

Dozens of dCs show  Balmer emission lines, signs of strong chromospheric activity. This is surprising, since activity is rare in older stars.  Activity is also associated with enhanced X-ray emission, which for dC stars remains uncharacterized.  Stellar activity - which correlates with rapid rotation in the general population (see \S\,\ref{sec:rot} below) - may be a result of spin-up by accretion in dC stars, {\, a hypothesis we aim to test here via X-ray observations}.  

\subsection{The Planetary Nebula Connection}
\label{sec:pne}

Once the hot core of an AGB star is revealed, it may illuminate the layers of expelled material to form a spectacular planetary nebula (PN). 
%\sout{A significant fraction of PNe are the products of interacting binary star systems (e.g., \citealt{DeMarco09,  Miszalski09}), which can influence the PN shape via formation of an accretion disk around the secondary or by angular momentum injection during, e.g., a common envelope phase.}
% and the  consequent generation of a disk and/or a strong magnetic
% dynamo at the primary.  
Binary companions to the central stars of PNe (CSPNe) are widely believed to be responsible for bipolar (axisymmetric) structures in PNe (see, e.g., \citealt{Balick02,Balick04,Garcia18}). 
%\sout{and necessary for outflow collimation and jet formation (e.g., \citealt{Boffin12, Corradi11, Blackman01}).}
% The complex and often asymmetrical PNe illuminated by their central stars (CSPNe) are now suspected to be strongly shaped by the presence of a binary companion (e.g., \citealt{Balick02,Balick04,Garcia18}.
% for refs, see Kastner 2012, AJ 144, 58
Many dC stars were likely born within planetary nebulae.  Indeed, the CSPN of PN~G054.2-03.4 (``The Necklace"; \citealt{Miszalski13a}) reveals a dC spectrum when the CSPN is in eclipse, offering {\em direct} evidence for this link between PNe and dC stars. 
% \sout{Several Ba\,II CSPNe are also known to exist (e.g., LoTr~5, LoTr~1, Hen2-39; \citealt{Bond03,Miszalski12,Miszalski13b,Tyndall14}).   Some are observed to be fast rotators \citep{Bond18}, further strengthening the connection of dCs and Ba\,II stars as close cousins in the family of post mass transfer binary systems.}

Surviving main sequence companions to CSPNe may be ``born-again'' -
resembling pre-main sequence stars, but with rejuvenated coronae and
(hence) luminous X-ray emission (\citealt{Soker02}).
{\em Chandra} imaging of CSPNe \citep{Montez10} has revealed that most of
the pointlike X-ray sources are too hard to be modeled as
blackbody emission from a pre-WD stellar photosphere ($\sim100
- 200$kK).  Among CSPNe with hard X-ray emission
($T_X\gax>\,6$MK), at least 75\% (6 out of 8) have known close or
rapidly rotating companions (\citealt{Kastner12, Montez15}). 
The companions are expected to remain active much longer
($>$Gyr; \citealt{West08}) than characteristic PN lifetimes
($\sim\,10^5$\,yr; e.g., \citealt{Frew08}).  
While the nearest known binary CSPNe with hard X-rays are at $\sim500$\,pc (NGC~1514 at $460\pm 8$\,pc and LoTr~5 at $499\pm12$\,pc; \citealt{Bailer-Jones18}), field dCs are {\, worthy of study because some are} as close as $\sim\,80\,$pc, and uncontaminated in the X-rays by hot CSPN radiation.

% In the Necklace, the observed kinematical age of the jets is twice the age of the ring in the Necklace, implying the jet and its polar caps formed {\em before} the CE phase, during mass transfer onto the secondary. The dC has surely been spun-up while gaining C-enriched mass ($\Delta M_2=0.03-0.35$ for an initial secondary mass $M_2= 1.0 - 0.4\,\msun$, respectively) from the accretion disk. As the newly-minted dC settles back to the main sequence from a thermally unstable phase (with radius potentially $\sim2.5\times$ greater than a normal dC (Af{\c s}ar \& Ibano{\v g}lu 2008), we expect a long-lasting ($\sim$Gyrs) active phase as the compact core cools and fades, a scenario we propose to test with nearby dC stars.  

\subsection{Binary Evolution Scenarios}

Mass transferred from the former AGB companion may occur via stellar wind, Roche-lobe overflow (RLOF), and/or common-envelope (CE) evolution. 
Models for dC formation in both the disk and halo \citep{deKool95} predict a bimodal orbital period distribution, with a large peak near a decade, for objects that have accreted AGB wind material with no substantial decrease of the orbital separation.  A smaller peak near 1 year likely contains systems that underwent a CE phase, where the companion was subsumed in the expanding atmosphere of the AGB star when it filled its Roche lobe.  These models reproduce the well-studied distributions of CH and Ba\,II giants, whose progenitors are almost certainly the dC stars.  However, many gaps remain in our understanding. 
For instance, using a variety of reasonable typical assumptions about the initial period distribution and the physics of the mass-transfer process, the synthetic populations of \citet{Abate18} under-predict the  frequency of short period systems relative to an unbiased sample of observed CEMP-$s$ stars  (e.g., \citealt{Iaconi18}).  
% \sout{Population syntheses are subject to the assumption of a CE $\alpha$ formalism that is known not to be easily applicable (e.g., \citealt{Wilson19}). Post-CE systems with binary orbital periods in excess of weeks are very difficult to reproduce in simulations (e.g., \citealt{Iaconi18}).  It is possible that a mass ratio close to unity at the time of CE may prolong the phase of Roche lobe overflow, resulting in longer periods as suggested by \citet{Reichardt19}, but not yet born out by simulations.}

Very small initial orbital separations may not readily result in
dC stars, because the primary will fill its Roche lobe, truncating evolution before the TP-AGB phase.  
% Furthermore, a CE phase that results in a dC must occur near the end of the primary's evolution for there to be sufficient carbon overabundance.  By then, the primary envelope mass is already substantially reduced, and the CE phase does not result in extreme loss of orbital angular momentum.  For this reason, very short dC orbital periods are not expected.  
However, orbits can evolve in a variety of ways as described e.g., by \citet{Chen18}.   Recent results from radial velocity monitoring of dCs \citep{Whitehouse18, Roulston19} find some dCs with unexpectedly short periods (1.2 days \citealt{Corradi11, Miszalski13a}; $\sim 3$ days; \citealt{Margon18}) or show large RV variations with as-yet undetermined orbits \citep{Roulston19}.
% \sout{Many white dwarf + main sequence (WDMS) systems have been discovered from SDSS spectra with M dwarf companions \citep{Rebassa12,Rebassa16} and some of them are clearly post common envelope binary (PCEB) systems, as painstakingly determined from photometric and/or radial velocity variability.  Most such PCEBs observed to date have periods shorter than 10d \citep{Jones17}.  However, there is considerable observational bias against detecting longer period systems \citep{DeMarco08, Jones17}, due in large part to reliance on existing photometry to find binaries.   Therefore, the evolutionary paths that lead to post-AGB binaries with periods 100--1000d, or their role in the formation and  morphology of PNe are still not well-known \citep{VanWinckel09,Oomen18}. Longer period binaries associated with the central stars of PNe (CSPNe) are now being found by radial velocity surveys (e.g., \citealt{VanWinckel14, Jones17, Miszalski18}). }
The dCs may prove quite important for further studies, because we know directly from their C$_2$ and CN bands that they are post-AGB mass transfer binaries, which makes them especially valuable laboratories of stellar binary evolution. 
 
\subsection{Stellar Rotation and Activity}
\label{sec:rot}

Since dC stars seem to be from older (thick disk or halo) populations \citep{Green13,Farihi18}, their congenital rotation rates, dynamo
strengths and related activity should have wound down significantly (e.g., \citealt{Gondoin18} and references therein).  However, \citet{Jeffries96} describe how the accretion of a massive, slow (10 - 20\,km/s) AGB wind is expected to spin up its low-mass secondary to short  ($\lax$10\,hr) rotation periods, for final orbital separations of about 100\,AU or less.  We may therefore expect significant activity and X-ray emission to be characteristic of dC stars - at least those in similar orbits - even if their progenitors  may be older than most active stars in the Galaxy.  

The activity lifetimes inferred for M dwarfs range from $\sim\,1 $ -- $5$\,Gyr (for M1 -- M4 stars; \citealt{West08}). 
The dC stars could remain active for similar timespans after mass transfer. The active fraction among late-type
stars as a function of age, metallicity and binary orbital elements is a current topic of strong interest (e.g., \citealt{West08,Morgan12}), but the strength, frequency or duration of activity in {\em rejuvenated} (post mass transfer) dwarfs remains unknown.  The dC stars offer a unique tracer, because even without measured orbital signatures of binarity,  their C$_2$\, and CN bands mark them clearly as post mass transfer systems.  

% STRUCTURE and RADIUS
Depending on the host system metallicity and
evolution, a dC star may have inherited a substantial fraction of its final mass by accretion (e.g., \citealt{Miszalski13a}), 
potentially changing its overall structure in the process. Assuming the accreting star is fully convective, then it must inherit 0.1 -- 0.2 \Msun\, of C/O $\sim $ 2--3 material.  By contrast a much smaller quantity of more highly-enriched (e.g., C/O$\,> 20$) material would suffice.  As mentioned above, the metallicity of the accretor is also key; for a star like G77-61 with extremely low metallicity \citep{Plez05}, much less mass transfer is required from a donor with the same C/O.

After a major mass accretion event, the dwarf's radius may remain inflated by accretion shocks, or even by enhanced magnetic activity itself, as debated for CSPNe (e.g., \citealt{Jones15}).   
% \sout{ Active stars % (including low-mass eclipsing binaries) are cooler and have slightly ($\sim$10\%) larger radii than inactive stars of similar mass (\citealt{Mullan01, Morales08}).  This could aid and prolong a phase where C$_2$ molecules can form.  Determining whether dC stars are inflated is difficult because to date, there are no published model atmospheres for dCs.  With measured parallax and photometry, we can estimate their bolometric luminosity (see \S\,\ref{sec:Lbol}),  but we have no way of knowing the expected equilibrium luminosity (and hence radius) for a given mass.}  
Nevertheless, long after the newly-minted dC settles back to the main sequence from a thermally unstable phase (with radius potentially $\sim2.5\times$ greater than a normal dC;  \citealt{Afsar08}), we expect a long-lasting ($\sim$Gyrs) active phase as the compact core cools and fades.  

% The unique properties of dC stars can help constrain the transferred amount of abundance-enhanced mass, how mixing occurs in the recipient dC, and spin-down times.  The dC stars thus provide a fresh avenue to understanding mass transfer in binaries - arguably the largest remaining uncertainty in stellar evolution modeling, with important implications for every class of interacting binaries (e.g., SN Ia, GRBs, CVs, and novae).  Study of a dC sample can also flesh out the latter chapters of the emerging story of the most freshly-minted dC stars, those in planetary nebulae.

Rapid rotation rates are typically found in young stars, as long as they retain  angular momentum from their initial collapse.  Rapid rotation together with convection is thought to drive an internal magnetic dynamo and consequent magnetic activity (e.g., \citealt{Kosovichev13} and references therein). The tangling, breaking and reconnection of magnetic field lines, as seen in the Sun, in turn generates chromospheric activity (associated observationally e.g., with H$\alpha$ emission) and coronal activity (associated with X-ray emission).  These tracers of magnetic activity decrease with stellar rotation rates (e.g., \citealt{Pallavicini81, Reiners12, Wright11}).
Rotation periods of main sequence stars with outer convection zones increase with age due to angular momentum loss through magnetized winds (e.g., \citealt{Kraft67, Matt15, Garraffo18}).

Even though we do not yet have rotation periods measured for  dCs, the detection of enhanced X-ray emission would provide very suggestive evidence that they have active dynamos, consistent with rotation rates enhanced by spin up from the accretion of angular momentum during past episodes of post-AGB binary mass transfer.  This possibility motivated the pilot \Chandra\, X-ray study whose results we describe below.

\section{Sample Selection}\label{sec:sample}

The list of six targets for this initial \Chandra\, study of X-rays from dC stars was compiled from our uniformly-selected SDSS sample of high latitude C stars \citep{Green13}.  We selected only those that are definitively main sequence, based on high proper motions measured between USNO-B and SDSS \citep{Munn04}.  (Any giant with such high proper motion would be nearby and much too bright for SDSS.) For the most resource-efficient exploration of dC stars, we restricted the sample to either DA+dC systems or dCs showing H$\alpha$ emission (dCe stars, henceforth) with $i<$17, yielding 6 objects (2 DA+dCs, and 4 dCes).  One was already observed in X-rays; SDSS~J125017.90+252427.6 falls 
% 192.574524 25.407665
serendipitously in both \XMM\, and \Chandra\, fields (\citealt{Green13}), as it is only about 6\,arcmin away from the galaxy NGC\,4725. 

\begin{figure}[ht]
%\plotone{compA.pdf}sroperties in the 0.3-3.0 keV energy range, which includes all detected photons.   X-ray source properties are shown in Table\,\ref{xobs}.  
\includegraphics[angle=270, scale=0.65]{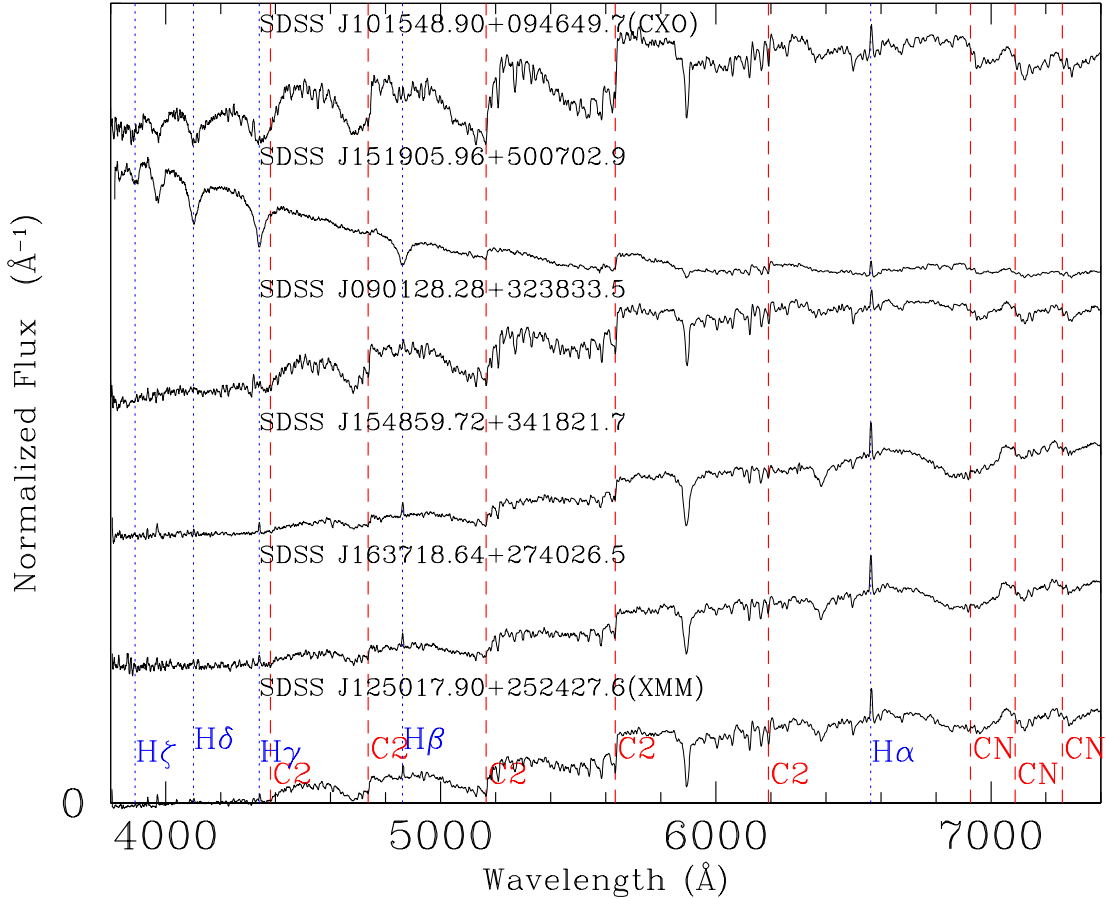}
\caption{\small
SDSS spectra of our full dC star sample.  All show strong  C$_2$  bands at $\lambda$4737, 5165, and 5636, as well as H$\alpha$ emission betraying chromospheric activity.  Broad Balmer absorption lines in the top 2 spectra reveal their hot white dwarf companions. C$_2$ bandhead positions are shown with red long-dashed lines, and Balmer line wavelengths with blue short-dashed lines. }
\label{fig:spec}
\vspace{0.5cm}
\end{figure}

The small pilot sample we consider here likely includes dC stars (1) with a relatively short time since accretion (the dC systems including still-hot DA white dwarfs) and/or (2) those still showing clear signs of activity (the dCe stars, for which the presumed DA white dwarf has cooled beyond detectability even in the optical/UV).  SDSS spectra are shown in Fig\,\ref{fig:spec} for the objects in our 
{\em Chandra} sample, illustrating the range of dC types herein. 

\begin{figure}[t!]
%\hspace*{-0.5cm}\includegraphics[angle=0,scale=0.28]{CMD1.png}
\plotone{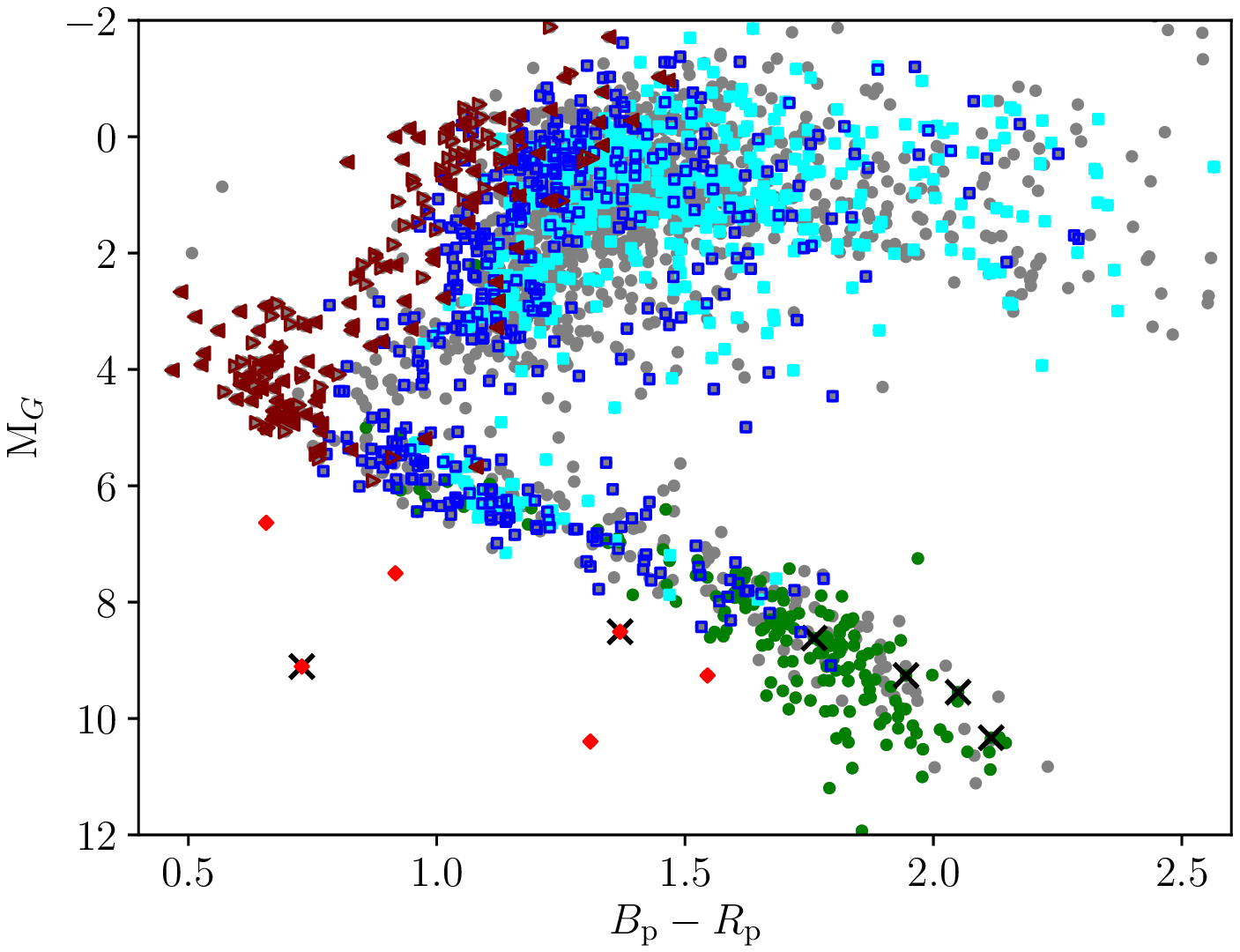}
\caption{A color-magnitude diagram derived from our compilation
  of the latest spectroscopic samples of carbon-enhanced stars,
  using Gaia DR2 to derive an absolute $G$-band magnitude, and showing
  only significant parallaxes, with $\varpi/\sigma_{\varpi}>5$.
  %parallax 5$\times$ its error.  
   Dwarf carbon (dC) stars, largely from SDSS \citep{Green13, Si14} and LAMOST \citep{Yoon16, Li18} form the main sequence,  ($6\lax M_G \lax 11$), most showing strong C$_2$ and CN molecular  bands.  Dwarfs identified in \citet{Green13} are shown by filled green circles.  Red diamonds mark known DA/dC systems.  The dCs observed with \Chandra\, and analyzed herein are marked with large black crosses. 
  CH (empty blue squares) and Ba\,II (filled cyan squares) from \citet{Li18}, and CEMP stars (maroon triangles) from \citet{Yoon16} are also shown. 
   %There is no significant separation evident between CEMP-$s$ (filled leftward maroon triangles) and CEMP-no stars (empty rightward maroon triangles).  
    As innocent bystanders to mass transfer, dCs are seen across a range of absolute magnitudes and temperatures, as long as they remain cool enough for C$_2$ and/or CN molecules to form \citep{Green13}.
    Most of the CH, Ba and CEMP stars are brighter than the turnoff
    ($M_G<5$), and may well be evolved dC stars. We have not corrected
    for reddening on this plot; interstellar reddening is
    insignificant for most dCs.  However, the reddest giants
    ($B_p-R_p\gax 1.8$) are at low galactic latitude, and likely
    significantly reddened.  A cross-correlation with the \Chandra\,
    CSC2.0 \citep{Evans10} and \XMM\, 3XMM-DR8 \citep{Rosen16}
    catalogs reveals
    no new matches among these stars.}
\label{fig:cmd}
\end{figure}

 For context, we make use of the distance information derived from Gaia Data Release 2 \citep{Gaia18} by \citet{Bailer-Jones18} to plot the range of color and absolute magnitude for a compilation of non-AGB carbon-enhanced stars in Fig\,\ref{fig:cmd}.  While originally selected based on their large proper motions, the dC systems in our sample are confirmed as definitively main sequence, having M$_G >8$.  

%\vspace{-0.5cm}

\section{X-ray Observations and Analysis}\label{sec:xrays}

Our targeted \Chandra\, observations of dC stars were performed using the S3 (backside illuminated CCD) on ACIS-S between 2014 October and 2016 May (proposals 15200243 and 16200105; P. Green P.I.).  Exposure times ranged from 16 to 28\,ksec per star.  
Some of the observations were split into several exposures (ObsIDs) as part of the requirement to keep various \Chandra\, susbsytem temperatures within their acceptable ranges. 
None of the observations uses a grating.  Most are in VFAINT ACIS mode,
but several are in FAINT mode; observation details are listed in
Table\,\ref{tab:xobs}.   

For SDSS J163718.64+274026.5, we analyzed the
merged observations, since they use the same instrument, and are only
separated by two days (and just 4$^{\circ}$ in roll angle).
SDSS~J125017.90+252427.6 falls within 3 archival observations, with
significantly different dates and instrument configurations, so we
chose to analyze the longest observation only (29.6\,ksec; PI Garmire). 

We reprocessed the \Chandra\, event lists with the CIAO (ver 4.10) \verb+chandra_repro+ script and CALDB (4.7.8), which account for afterglows, bad pixels, charge transfer inefficiency, and time-dependent gain corrections.
All our dCs were detected in every \Chandra\, ObsID, at positions
%**
within 2\,\arcsec\, of those measured in Gaia DR2 (epoch 2015.5).
We used the CIAO \verb+srcflux+ tool to estimate source properties in the 0.3-3.0 keV energy range, which includes all detected photons.  X-ray source properties are shown in Table\,\ref{tab:fluxes}.  
To estimate
intervening absorption from extinction due to Milky Way dust, we used the 
\texttt{dustmaps} Python package (Bayestar17) of \citep{Green18a}; for every dC direction, these columns were inconsequential. 

For each dC, we derived two X-ray flux estimates, based on either a 2\,MK or a 10\,MK
optically-thin plasma (APEC; \citealt{Smith01}) with absorption modeled using WABS \citep{Morrison83}.  As described in \S\,\ref{xmodels}, our motivation for these choices comes both from the literature, and from attempts to fit the \Chandra\, spectra of the two dCs herein with the most counts.   Further discussion of our sample below includes results assuming either 2 or 10\,MK plasma temperatures.

%\begin{figure}[h]
%\includegraphics{dCStars_J1519_spec.eps}{dCStars_J1519_cont.eps}
%\caption{\small
%\Chandra\, spectrum of the DA/dC SDSS J151905.90+500702.9 }
%\label{fig:xspec}
%\end{figure}

\begin{deluxetable*}{lcccccccc}[ht]
\tablecaption{{\em Chandra} X-ray Observations \label{tab:xobs}}
\tablecolumns{6}
\tablewidth{0pt}
\tablehead{
\colhead{Object} &
\colhead{Type} &
\colhead{ObsID} &
\colhead{Obs-Date} & \colhead{Exposure} & \colhead{Inst.}  & \colhead{Mode} %\\
%& & \colhead{(YYYY-mm-dd)} & \colhead{(ks)} &
\\
 & & & & \colhead{(ksec)} 
}
\startdata
SDSS J090128.28+323833.5 & dCe  & 16650 & 2015-12-12 & 18.3 & ACIS-S & VF \\
SDSS J101548.90+094649.7 & DA/dC & 15702 & 2015-01-18 & 15.9 & ACIS-S & F \\
SDSS J125017.9+252427.6  & dCe  & 409 & 1999-12-20& 17.5 & ACIS-S & F \\
                         &      & 2976 & 2002-12-02 & 24.6 & ACIS-S & F \\
                         &      & 17461 & 2016-05-09 & 29.5 & ACIS-I & VF \\
SDSS J151905.90+500702.9 & DA/dC & 16649 & 2016-05-16 & 27.6 & ACIS-S & VF \\
SDSS J154859.72+341821.7 & dCe   & 16651 & 2015-06-17 & 16.0 & ACIS-S & VF \\
SDSS J163718.64+274026.5 & dCe & 17534 & 2014-10-19 & 8.5 & ACIS-S & VF \\
 &  & 16652 & 2014-11-21 & 17.2 & ACIS-S & VF \\
\enddata
%\tablenotetext{a}{}
%\tablecomments{}
\end{deluxetable*}

% We might consider evaluating dCs generally for the presence of dust using e.g., WISE, but again, compared to what stellar models? 

Our X-ray analysis is very unlikely to be complicated by emission from 
the white dwarfs, even in the DA+dC systems, because hot white 
dwarf photospheric X-ray emission is extremely soft.   For
 white dwarfs in the field, emission of hard X-rays is almost always 
 attributed to coronal emission from a late-type dwarf companion
 (\citealt{O'Dwyer03}). Indeed, among the thousands of WDs that have been observed (mostly serendipitously) by \XMM\, or ROSAT, only a handful of single WDs have spectra that are hard like the CSPN (\citealt{Bilikova10}).  The great majority of the CSPNe with hard X-ray emission that do not have an easily-detected late-type companion may have e.g., an accreting white dwarf companion \citep{Miszalski19}.

Hydrogen Balmer line emission from the stellar chromosphere is a known tracer of activity; H$\alpha$ line luminosity is well-correlated with projected rotation velocity $v$\,sin$i$ (e.g., \citealt{Maldonado17} for M0 - M4 dwarfs) after excluding spectroscopic binaries.   While all of the dCs in our sample have H$\alpha$ emission evident by selection, we are unable to reliably use the H$\alpha$ equivalent widths in these dCs to predict their X-ray emission. H$\alpha$ equivalent widths are not a suitable indicator of stellar activity and must be transformed into H$\alpha$ fluxes, as found by multiple studies \citep{Reiners12,  Maldonado17}, before fitting H$\alpha$ - X-ray relations.  However the correlation between H$\alpha$ flux and X-ray flux appears to be a weak one even in K and M dwarfs, and becomes only more problematic with dCs given the lack of published model atmospheres needed to transform H$\alpha$ equivalent width into H$\alpha$ flux. We therefore did not attempt to use the observed H$\alpha$ emission to predict X-ray fluxes.

\section{Bolometric Luminosities}
\label{sec:Lbol}

Bolometric luminosity estimates are important for placing dCs in the context of other active stars, because stellar activity is best characterized in the X-rays across a range of stellar types when the X-ray luminosity is normalized by the bolometeric luminosity and $\Lbol$ can vary by a factor of $\sim$20 across the range of main sequence spectral types known to show C$_2$ and CN molecular bands.  To estimate bolometric luminosities $\Lbol$, we assembled a spectral energy distribution (SED) for each dC star in our sample from public
cataloged photometry and calculated \Lbol\ using the \texttt{sedkit} Python package \citep[][Filippazzo et al., in prep]{Fili15}. The exact procedure is detailed in those works and summarized here.

Optical, near-infrared, and mid-infrared magnitudes were obtained from
the SDSS Data Release 12 \citep{Alam15}, 2MASS Point Source Catalog
\citep{Skru06}, and AllWISE Source Catalog \citep{Cutr13},
respectively. We converted photometry to the VegaMag
system and corrected for (small) interstellar extinction using the
\texttt{dustmaps} Python package \citep{Green18a}. Synthetic
magnitudes were calculated from the SDSS spectrum, which was then
normalized to the observational photometry. In wavelength regions with
no spectral coverage, the SED was linearly interpolated between
photometric points.  For the DA/dCs, we excluded the SDSS $u, g$ and $r$ band 
photometry and spectral flux, since they are
contaminated by the DA white dwarfs. To approximate a Wien tail, the SED was linearly interpolated from the shortest wavelength data point down to zero flux at zero wavelength. A blackbody spectrum fit to the WISE photometry was used to approximate a Rayleigh-Jeans tail at long wavelengths. We then calculated the bolometric flux \fbol\ as the integral of the complete SED.
%(Tables \ref{tab:opt-phot} and \ref{tab:ir-phot}) removed as unecessary by PG

Finally, we calculated \Lbol = 4$\pi$\fbol d$^2$ for each source using
a parallax measurement from the Gaia DR2 and
the resulting distances from \citet{Bailer-Jones18}. The results are
shown in Table \ref{tab:distLbol}, where we adopt the solar bolometric
luminosity value of log\,$L_{\odot}=33.583$.

% ***check GALEX too

\begin{deluxetable*}{lcccc}
\tablecaption{Parallaxes, Distances and Bolometric Luminosities \label{tab:distLbol}}
\tablecolumns{5}
\tablewidth{0pt}
\tablehead{
\colhead{Object} &
\colhead{$\varpi$} &
\colhead{Distance} &
\colhead{log (\Lbol/$L_\odot$)}
\\
 & \colhead{(mas)} & (pc) & erg/s
}
\startdata
%J0901&1.7083 $\pm$ 0.1360&-1.292 $\pm$ 0.069\\  Nh= 1.6\pm 1.05e20
%J1015&2.1046 $\pm$ 0.1264&-1.357 $\pm$ 0.053\\  2.6\pm 1.05
%J1250&3.5374 $\pm$ 0.0793&-1.519 $\pm$ 0.020\\  1.05\pm 1.05
%J1519&2.2568 $\pm$ 0.0773&-1.464 $\pm$ 0.030\\  1.05\pm 1.05
%J1548&4.3237 $\pm$ 0.0806&-1.867 $\pm$ 0.017\\  1.05\pm 1.05
%J1637&2.4664 $\pm$ 0.0928&-1.622 $\pm$ 0.034\\  3.14\pm 1.57
% PG repaired sigfigs added Bailer-Jones distances, with error and average of r_lo and r_hi
% These are numbers from Filippazzo/Notebook2018jul19/
% scaled just slightly by me to Bailer & Jones
J0901& 1.71 $\pm$ 0.14 & 577.1 $\pm$ 46.5  & $-1.26 \pm$ 0.07 \\ 
J1015& 2.10 $\pm$ 0.13 & 469.9 $\pm$ 28.6  & $-1.32 \pm$ 0.05 \\
J1250& 3.54 $\pm$ 0.08 & 280.6 $\pm$  6.3  & $-1.52 \pm$ 0.02 \\
J1519& 2.26 $\pm$ 0.08 & 438.0 $\pm$ 15.0  & $-1.47 \pm$ 0.03 \\
J1548& 4.32 $\pm$ 0.08 & 229.9 $\pm$  4.3  & $-1.87 \pm$ 0.07 \\
J1637& 2.47 $\pm$ 0.09 & 401.6 $\pm$ 15.1  & $-1.56 \pm$ 0.03 \\
\enddata
%\tablenotetext{a}{}
%\tablecomments{}
\end{deluxetable*}

% ***
%Analysis of the \XMM\, spectrum of SDSS~J125017.90+252427.6 suggests coronal-like X-ray emission from a hot ($k$T$\sim$0.25$\pm0.07$\,keV) optically thin plasma with very little intervening absorption ($N_H\sim10^{20}$\atoms) for an intrinsic X-ray flux of $7.2\times10^{-15}\,\fcgs$.  At its distance of about 280\,pc and assuming $\Lbol\sim 1.6\times10^{32}$\,erg/s this yields log\,$L_X/\Lbol \sim -3.3$.  
% Lx/LBol = 4*pi* ( 288*pc )^2 * 7.2e-15/1.6e32 = 0.00044659 

\begin{deluxetable*}{lccccccccc}[b!]
\tablecaption{X-ray Source Properties in the 0.3-3.0 keV Energy Range \label{tab:fluxes}}
\tablecolumns{7}
\tablewidth{0pt}
\tablehead{
\colhead{Object} &
\colhead{Chandra} & \colhead{Net CR} & \colhead{$T_{\rm X}$} & \colhead{$F_{\rm X,obs}$} &  \colhead{$F_{\rm X}$} &  \colhead{$L_{\rm X}$} \\
 & ObsID& \colhead{(cnt ks$^{-1}$)} & \colhead{(MK)} & %\colhead{($10^{-15}$ erg~cm$^{-2}$~s$^{-1}$)} & %\colhead{($10^{-15}$ erg~cm$^{-2}$~s$^{-1}$)} & 
 \multicolumn{2}{c}{$10^{-15}$ erg~cm$^{-2}$~s$^{-1}$)} &
 \colhead{($10^{28}$ erg s$^{-1}$)}
}
\startdata
J0901 & 16650 & 0.16$\pm$0.09 & 2 & $3.18^{+4.33}_{-2.25}$ & $3.77^{+5.14}_{-2.67}$ & $15.47^{+21.12}_{-11.02}$ \\
\ldots & \ldots & \ldots      & 10 & $0.97^{+1.33}_{-0.69}$ & $1.03^{+1.41}_{-0.73}$ & $4.23^{+5.79}_{-3.01}$ \\
J1015 & 15702 & 0.25$\pm$0.13 & 2 & $3.64^{+4.00}_{-2.30}$ & $4.82^{+5.28}_{-3.04}$ & $13.02^{+14.28}_{-8.25}$ \\
\ldots & \ldots & \ldots      & 10 & $1.34^{+1.48}_{-0.85}$ & $1.48^{+1.63}_{-0.93}$ & $4.00^{+4.41}_{-2.53}$ \\
J1250 & 409 & 1.62$\pm$0.99   & 2 & $20.40^{+29.00}_{-15.08}$ & $22.90^{+32.50}_{-16.93}$ & $21.91^{+31.09}_{-16.20}$ \\
\ldots & \ldots & \ldots      & 10 & $8.48^{+12.02}_{-6.27}$ & $8.83^{+12.57}_{-6.53}$ & $8.45^{+12.03}_{-6.25}$ \\
\ldots & 2976 & 0.64$\pm$0.17  & 2 & $12.80^{+6.50}_{-5.00}$ & $14.40^{+7.30}_{-5.64}$ & $13.78^{+6.99}_{-5.40}$ \\
\ldots & \ldots & \ldots      & 10 & $4.13^{+2.11}_{-1.61}$ & $4.31^{+2.19}_{-1.68}$ & $4.12^{+2.10}_{-1.61}$ \\
\ldots & 17461 & 0.41$\pm$0.12 & 2 & $47.30^{+27.00}_{-19.70}$ & $53.10^{+30.30}_{-22.10}$ & $50.80^{+29.01}_{-21.17}$ \\
\ldots & \ldots & \ldots      & 10 & $6.03^{+3.45}_{-2.51}$ & $6.28^{+3.59}_{-2.62}$ & $6.01^{+3.44}_{-2.51}$ \\
J1519 & 16649 & 0.54$\pm$0.14 & 2 & $12.80^{+6.30}_{-4.77}$ & $14.40^{+7.10}_{-5.39}$ & $33.83^{+16.72}_{-12.72}$ \\
\ldots & \ldots & \ldots      & 10 & $3.52^{+1.75}_{-1.31}$ & $3.67^{+1.81}_{-1.37}$ & $8.62^{+4.26}_{-3.23}$ \\
J1548 & 16651 & 0.31$\pm$0.14 & 2 & $5.35^{+5.15}_{-3.14}$ & $6.00^{+5.80}_{-3.52}$ & $3.84^{+3.71}_{-2.25}$ \\
\ldots & \ldots & \ldots      & 10 & $0.26^{+1.06}_{-0.26}$ & $0.27^{+1.10}_{-0.27}$ & $0.17^{+0.70}_{-0.17}$ \\
J1637 & 16652 & 0.57$\pm$0.18 & 2 & $7.90^{+5.00}_{-3.49}$ & $11.00^{+7.00}_{-4.85}$ & $21.65^{+13.80}_{-9.58}$ \\
\ldots & \ldots & \ldots      & 10 & $3.03^{+1.91}_{-1.34}$ & $3.40^{+2.15}_{-1.50}$ & $6.69^{+4.24}_{-2.96}$ \\
\ldots & 17534 & 1.05$\pm$0.35 & 2 & $14.20^{+9.50}_{-6.59}$ & $19.80^{+13.20}_{-9.20}$ & $38.97^{+26.02}_{-18.16}$ \\
\ldots & \ldots & \ldots      & 10 & $5.50^{+3.68}_{-2.55}$ & $6.18^{+4.12}_{-2.86}$ & $12.16^{+8.12}_{-5.65}$ \\
\enddata
%\tablenotetext{a}{}
%\tablecomments{}
\end{deluxetable*}

\section{Rotation-Activity Relationship}
\label{sec:rot-act}

%** do this (better) in the intro.  Need it twice?
Field main sequence stars with outer convection zones are magnetically active, as traced by coronal X-rays and/or chromospheric H$\alpha$ emission.   The large observed range in log\,\LxLbol\, from about 10$^{-3}$ down to 10$^{-8}$, is thought to occur because stars' rotation rates slow with age, due to mass loss from stellar winds (e.g., \citealt{Skumanich72}).  Indeed, rotation rates correlate strongly with activity, especially when normalized by the convective turnover time $\tau$ via the Rossby number $R_0=P_{rot}/\tau$ \citep{Noyes84}. The magnetic dynamo that drives these signatures of activity is thought to be generated by differential rotation inside the star.  At the highest rotation rates, corresponding to $R_0\lax 0.13$, activity saturates, and log\,\LxLbol\, remains at $\sim\,-3.3$ \citep{Micela85, Wright11}.   It was long thought that this correlation must be mediated by an $\alpha\Omega$ dynamo \citep{Parker55}, which requires an interface (the tachocline) between a solidly rotating radiative core, and a differentially rotating convective envelope.  However, the same rotation-activity relationship appears to hold even for late-type stars thought to be fully convective \citep{Wright18}.
% e.g., for stars later than types M3-–M4.  (B-−V$\sim$1.5-–1.7).

Unfortunately, we do not have the sensitive multi-epoch photometry
that would be required to measure the rotation rate for the dC stars
in our sample.  The greatest promise to achieve this in the near
future - at least for the brighter dC examples - is after NASA's Transiting Exoplanet Survey Satellite (TESS) satellite \citep{Ricker15} surveys the northern sky, providing photometric measurements every 30 minutes,
which can be combined as needed for greater sensitivity (at the cost
of lower time resolution). Neither do we have estimates for the convective turnover times, due to a paucity of stellar structure models for dC stars.  However, since we have empirical measurements of both $\Lx$ and $\Lbol$, we can estimate the range of likely values of $P_{rot}$ or $R_0$ for dC stars, reasonably assuming that their internal dynamics and magnetic dynamo have stabilized since the end of mass transfer.  

\begin{figure}[h!]
\plotone{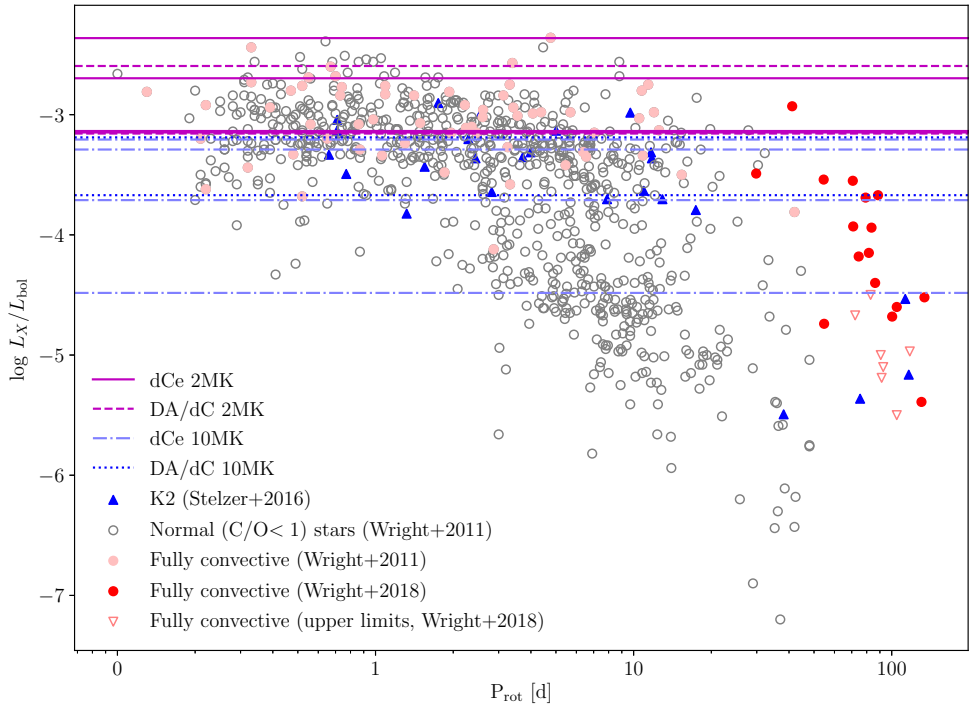}
\caption{\small
The activity level of dC stars in our sample in context with activity in normal (C/O$<1$) main sequence stars.  The logarithm of X-ray to bolometric luminosity ratio log\,$\LxLbol$ is plotted vs. rotation period for the normal dwarf stars in the sample of \citet{Wright11} (open black circles).  Fully convective dwarfs are highlighted from \citet{Wright18} (red circles, {\, or red triangles for the upper limits}) and from \citet{Wright11} (faint red circles).  Stars from the Kepler Two-Wheel (K2) samples of \citet{Stelzer16} are shown as blue triangles.  The horizontal magenta (blue) lines show the  log$\LxLbol$ values for the 6 dC stars observed by \Chandra\, in our pilot program, assuming a plasma temperature of $T_X=2MK$ ($T_X=10MK$).  Dashed (dotted) lines represent the 2 DA/dC systems (with a hot white dwarf still visible in the optical spectrum).  Solid (dot-dash) lines are for the other dC systems. 
%For bright dCs, $P_{rot}$ may be obtainable from TESS scans. 
}
\label{fig:ProtLxLb}
\end{figure}

In Fig\,\ref{fig:ProtLxLb}, We show the activity level of dC stars in our sample as a functino of rotation period, in context with activity in normal (C/O$<1$) main sequence stars from several recent sources in the literature.
\citet{Wright18} studied the coronal activity-rotation relationship for late-type stars with rotation periods known from the MEarth project \citep{Nutzman08}, including 
% ** They fit spectra to estimate fluxes, getting 0.3<kT<1keV or 3<T<10MK
main sequence stars thought to be fully convective - those later than about M3 ($\Teff\,\lax 3300$K), corresponding to masses below about $\sim 0.35\Msun$ \citep{Chabrier97}.   \cite{Stelzer16} compiled X-ray data from ROSAT and \XMM\, catalogs for K2 Superblink stars with well-measured periods.  They assumed a thermal (APEC) single temperature  model (APEC) of 3.5MK
 % kT = 0.3\,keV 
 and a column density of $10^{19}\atoms$. 

From Fig\,\ref{fig:ProtLxLb}, limits on the rotatino periods for these dC stars appear to be in the 'saturated' regime where log$\LxLbol\geq-3.3$, at least when assuming plasma temperatures of $\sim 2$\,MK.  About half remain in the saturated regime even when assuming a high plasma temperature $\sim$10\,MK.
% While rotation periods are not available for these dCs, 
If the lower temperature applies, then dCs' strong X-ray activity is consistent with rapid rotation rates of several days or less, as might be expected from accretion spin-up.  If the higher temperature applies, then periods are more weakly constrained, to about 20\,days or less.

We reiterate the caveats that (a) this pilot sample is not representative of dCs more generally since as mentioned, all 6 dC optical spectra in our sample show weak H$\alpha$ emission lines (see Fig\,\ref{fig:spec}) 
and (b) X-ray counts are too few to actually constrain $T_X$.
%  for reasonable higher assumed temperatures (e.g., $T\sim10MK$), several of these $\LxLbol$ values are no longer in the saturated 

\section{Summary}\label{sec:summary}

We have sought to test the hypothesis that dC stars, having inherited significant mass from an extinct AGB companion, could have spun up significantly, and that they may therefore show signs of coronal activity often associated with rapid rotation.  As a test for such coronal activity, we selected a small sample of dCs for X-ray observation with \Chandra.
% For a pilot \Chandra\, program, we have detected X-ray emission from all objects, albeit weakly, with log\,$L_X\sim 29$.  The small sample observed contains intriguing, but notably unrepresentative stars, selected precisely to achieve a high likelihood of detection;  all the objects show  H$\alpha$ emission, and two also have hot WD companions.
All 6 dC systems observed were detected by \Chandra, but with too few counts to allow useful spectral constraints.  We therefore calculate X-ray fluxes assuming plasma temperatures of 2MK and 10MK, bracketing a range of reasonable values.  We have estimated the dCs' bolometric luminosities directly by compiling and integrating multi-wavelength photometry, and using distances from Gaia DR2 parallaxes published after our \Chandra\, observations were performed. When assuming a reasonable plasma temperature of 2MK, all 6 systems appear to have log$\LxLbol\,\gax -3$, putting them in the saturated regime so that they might be expected to have short ($\lax$10d) rotation periods consistent with strong accretion spin-up.  At the upper range of expected temperature, near 10\,MK, the X-ray activity of dCs could imply longer ($\lax$20d) periods similar to less active dwarfs.

The X-ray emission from dC stars to date supports the mass transfer hypothesis.  Where dCs have been part of a central binary system of a PN, which could be common, then the activity we detect here argues for spin-down significantly longer than ($\sim\,10^4$\,year) PN timescales, because for most dC stars the former CSPNe are now WDs cooled beyond detectability.  

To better understand the characteristics of a more typical sample, sufficiently sensitive X-ray observations of the closest known dC stars with reliable distances from Gaia DR2 are critical.  Ideally, at least $\sim$100 counts per object should be obtained, sufficient to constrain individual plasma temperatures to within about $\sim$30\%.  The X-ray luminosity fractions $\LxLbol$ can then serve as a more reliable proxy to measure rotation and total accretion, within the context of models predicting longevity and intensity of the binary interaction during the AGB phase.  For instance, accretion of both mass and angular momentum is enhanced if there is a longer RLOF phase before the CE phase. A dC enriched by a prolonged period of RLOF would have relatively stronger X-ray emission, faster rotation, and perhaps also higher C/O.  Very close PCEB dC systems that have low X-ray activity may indicate a brief pre-CE phase. A dC that experienced only wind accretion may have low rotation and low $L_X$.  A significant sample of dCs with measured orbital and rotation periods and X-ray luminosities would greatly enhance our understanding of these useful and intriguing systems.

\vspace{1cm}
\acknowledgments

The scientific results reported in this article are based  on observations made
by the \Chandra\, X-ray Observatory, and data obtained from the \Chandra\,
Data Archive.  This research has made use of software provided by the \Chandra\, X-ray Center (CXC) in the application packages CIAO and Sherpa.

Support for this work was provided by the National Aeronautics and Space Administration through Chandra Award Numbers GO4-15005X and GO5-16004X issued by the Chandra X-ray Center, which is operated by the Smithsonian Astrophysical Observatory for and on behalf of the National Aeronautics Space Administration under contract NAS8-03060.  
BM acknowledges support from the National Research Foundation (NRF) of South Africa.

Funding for the Sloan Digital Sky Survey IV has been provided by the
Alfred P. Sloan Foundation, the U.S. Department of Energy Office of
Science, and the Participating Institutions. SDSS acknowledges
support and resources from the Center for High-Performance Computing at
the University of Utah. The SDSS web site is www.sdss.org.

SDSS is managed by the Astrophysical Research Consortium for the Participating Institutions of the SDSS Collaboration including the Brazilian Participation Group, the Carnegie Institution for Science, Carnegie Mellon University, the Chilean Participation Group, the French Participation Group, Harvard-Smithsonian Center for Astrophysics, Instituto de Astrof\'{i}sica de Canarias, The Johns Hopkins University, Kavli Institute for the Physics and Mathematics of the Universe (IPMU) / University of Tokyo, Lawrence Berkeley National Laboratory, Leibniz Institut f\"{u}r Astrophysik Potsdam (AIP), Max-Planck-Institut f\"{u}r Astronomie (MPIA Heidelberg), Max-Planck-Institut f\"{u}r Astrophysik (MPA Garching), Max-Planck-Institut f\"{u}r Extraterrestrische Physik (MPE), National Astronomical Observatories of China, New Mexico State University, New York University, University of Notre Dame, Observat\'{o}rio Nacional / MCTI, The Ohio State University, Pennsylvania State University, Shanghai Astronomical Observatory, United Kingdom Participation Group, Universidad Nacional Aut\'{o}noma de M\'{e}xico, University of Arizona, University of Colorado Boulder, University of Oxford, University of Portsmouth, University of Utah, University of Virginia, University of Washington, University of Wisconsin, Vanderbilt University, and Yale University.

\facility{2MASS, CXO, Gaia, Sloan, WISE}

\software{\texttt{Astropy} \citep{astropy18},  \texttt{matplotlib} \citep{matplotlib05}, \texttt{Numpy} \citep{NumPy06}}

\appendix
\section{X-ray Spectral Models}
\label{xmodels}

Our motivation for using 2MK and 10MK plasma temperatures for the X-ray spectral models of dC stars comes both from the literature and from our \Chandra\, observations.  The higher temperature model is consistent with median photon energies of $\sim 0.5 - 1.0$\,keV (6 -- 12 MK) seen in X-ray selected stellar samples observed with \Chandra\, from e.g., the COSMOS survey  \citep{Wright10}, and also with the temperature of peak emissivity for \ion{Mg}{11}.
However, for some single stars with detailed X-ray spectral fitting, plasma temperatures are close to 2\,MK (e.g., see the compilation in Table~9 of \citealt{Testa04}), the temperature of peak emissivity for \ion{O}{7}.

Even with very few counts detected in our \Chandra\, images, we can examine the median energy values and their location with respect to the thermal models and foreground extinction.  Most of the observed counts suggest $T_X > 3$\,MK and likely near 8--10\,MK, but this is uncertain without knowing what the circumstellar environment might contribute to extinction. 

% ***These are reasonable choices, based on multi-temperature fit results to stellar coronal emission e.g., from the COUP project (Feigelson...)??  No, those are all PMS stars, so probably not applicable.

\begin{figure}%
 \centering
 \subfloat[Counts spectrum]{{\includegraphics[width=8cm]{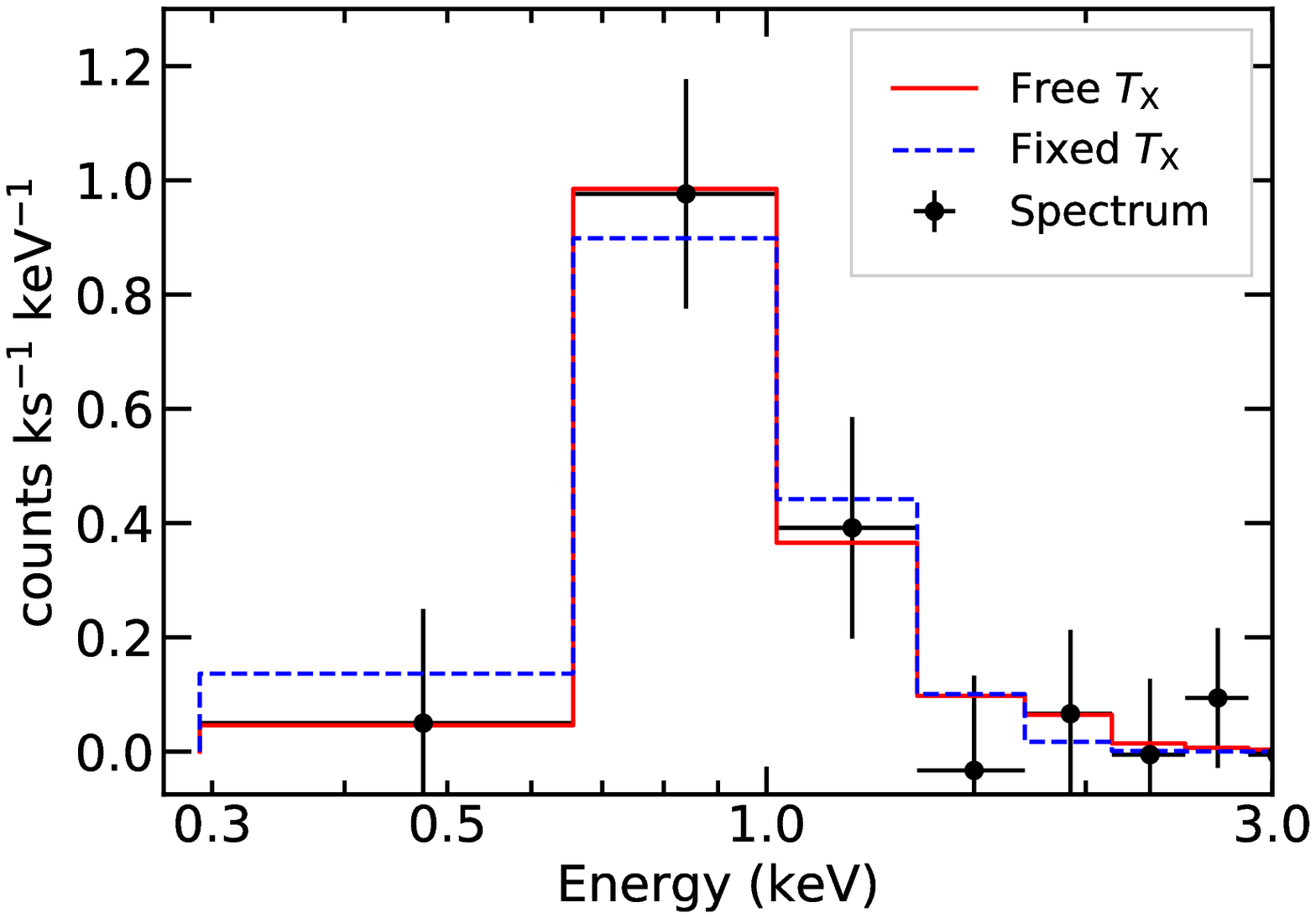} }}%
 \qquad
 \subfloat[Best-fit X-ray spectral parameters]{{\includegraphics[width=8cm]{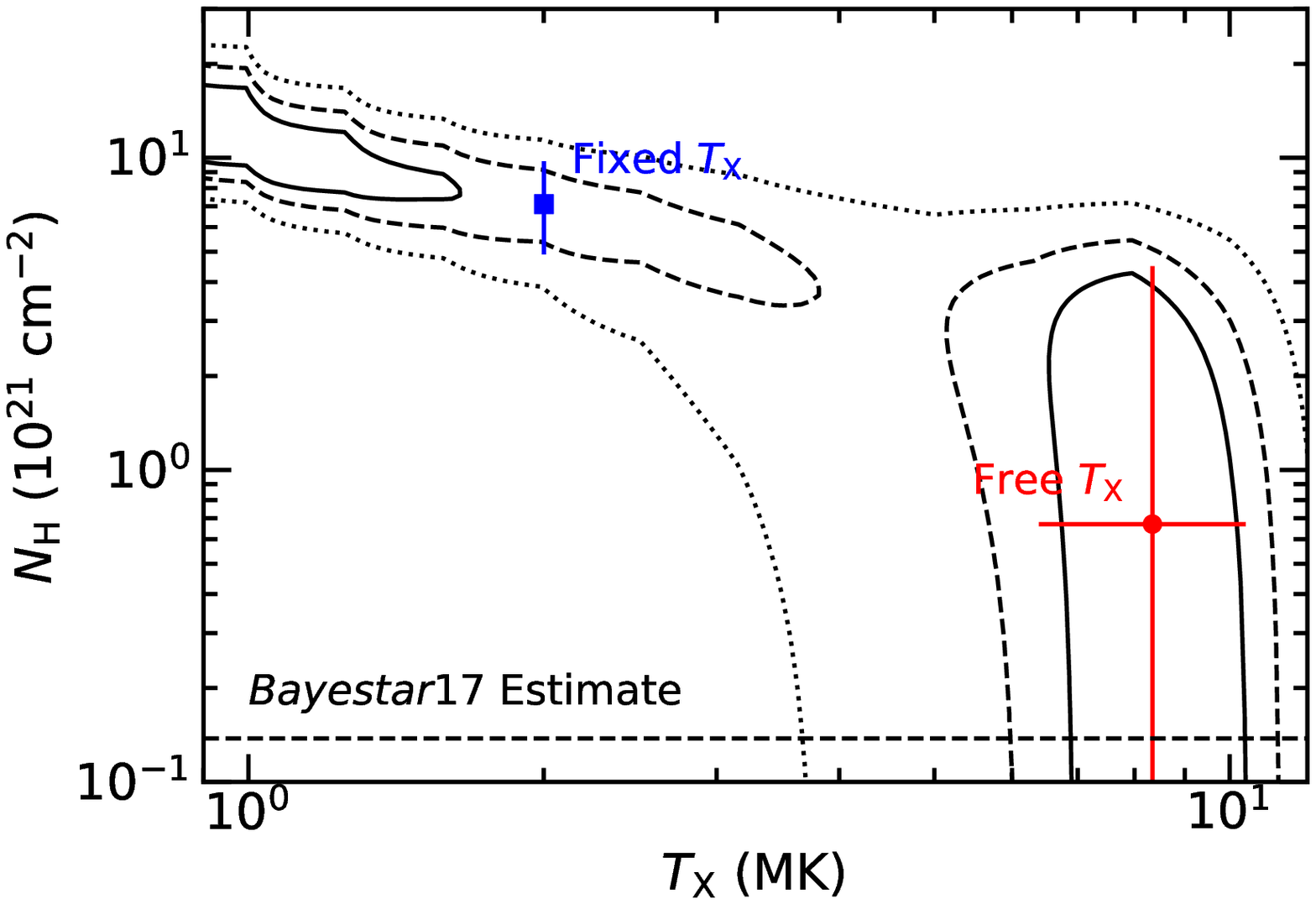} }}%
 \caption{\small  \Chandra\, spectrum of SDSS J151905.90+500702.9  from ObsID 16649.
(a) The photon spectrum in counts ksec$^{-1}$ keV$^{-1}$.  The expected counts for both 2\,MK and 8\,MK temperature plasma models are overplotted as blue dashed and red solid lines, respectively.     Since the total number of net counts is only $\sim$15, the fits are equally good, and quite poorly constrained.  (b) The best-fit contours (at 68.3, 90, and 99\%) for the APEC spectral fit parameters of temperature $T_X$ and intervening hydrogen column $N_H$.   The intervening column estimated from \citet[{\it Bayestar17};][]{Green18a} is quite low (dashed horizontal line), and is consistent with the best-fit spectrum for plasma temperature near $\sim$8\,MK. If we fix the temperature at $T_X=2$\,MK, a larger column  is required, which may be consistent with possible circumstellar material. 
 }
\label{fig:xspec15}
\end{figure}

\begin{figure}%
 \centering
 \subfloat[Counts spectrum]{{\includegraphics[width=8cm]{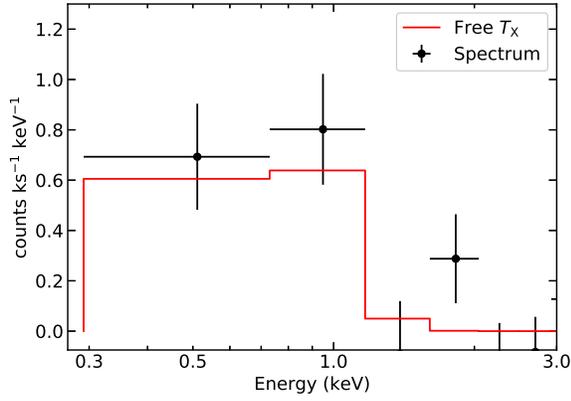} }}%
 \qquad
 \subfloat[Best-fit X-ray spectral parameters]{{\includegraphics[width=8cm]{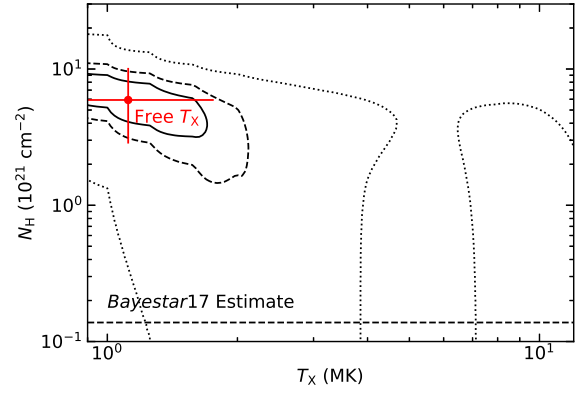} }}%
 \caption{\small  \Chandra\, spectrum of SDSS J125017.9+252427.6 from ObsID 2976.
(a) The photon spectrum in counts ksec$^{-1}$ keV$^{-1}$.  The expected counts for the best-fit $\sim\,1$MK APEC model is overplotted in red.       (b) As in Fig\,\ref{fig:xspec15}, the APEC spectral fit parameter contours of temperature $T_X$ and intervening hydrogen column $N_H$. The best-fit temperature is near $\sim 1$MK, but requires an intervening column density much larger than expected from the ISM, broadly consistent with possible circumstellar material. 
 }
\label{fig:xspec12}
\end{figure}

We can also analyze the two dCs in our sample with the largest number of counts, which are nonetheless insufficient for strong spectral model constraints.
SDSS~J151905.90+500702.9 (Fig\,\ref{fig:xspec15})  indicates that either a hot plasma temperature ($\sim 10$~MK with low absorbing column, $N_{\rm H}<5\times10^{21} {\rm ~cm}^{-2}$) or a cool plasma temperature ($\lax 2$~MK, with high absorbing column, $N_{\rm H}>5\times10^{21} {\rm ~cm}^{-2}$) might be applicable. 
The best-fit spectral model for SDSS~J125017.9+252427.6  (Fig\,\ref{fig:xspec12}) is single-peaked towards a cool value near $\sim$1\,MK.  

These crude spectral fits further justify our adopted spectral models of 2 and 10\,MK.  Formally they may allow for the possibility of significant circumstellar columns above the expected intervening values, but the constraints are very weak. Cool post-AGB wind material may indeed surround dCs, but such an intriguing possibility must be investigated with much more sensitive X-ray and/or sub-millimeter observations (e.g., \citealt{Bujarrabal13}) before being taken seriously.

\clearpage
\bibliographystyle{yahapj}

\begin{thebibliography}{}
% \bibitem[Aaronson et al.(1982)]{Aaronson82}Aaronson, M., Liebert, J., \& Stocke, J. 1982, \apj, 254, 507 
% \bibitem[Aaronson et al.(1983)]{Aaronson83}Aaronson, M., Hodge, P.~W., \& Olszewski, E.~W.\ 1983, \apj, 267, 271
\bibitem[Abate et al.(2018)]{Abate18} Abate, C., Pols, O.~R., \& Stancliffe, R.~J.\ 2018, \aap, 620, A63.
\bibitem[Af{\c{s}}ar \& Ibano{\v{g}}lu(2008)]{Afsar08} Af{\c{s}}ar, M., \& Ibano{\v{g}}lu, C.\ 2008, \mnras, 391, 802.
% \bibitem[Ag{\"u}eros et al.(2009)]{Agueros09}Ag{\"u}eros, M.~A., Anderson, S.~F., Covey, K.~R., et al.\ 2009, \apjs, 181, 444 
\bibitem[Alam et al.(2015)]{Alam15} Alam, S. et al. \ 2015, The Astrophysical Journal Supplement Series, 219, 12 
% \bibitem[Alksnis et al.(2001)]{Alksnis01} Alksnis, A., Balklavs, A., Dzervitis, U., et al.\ 2001, VizieR Online Data Catalog, 3227, 0 
\bibitem[Astropy Collaboration et al.(2018)]{astropy18} Astropy Collaboration, Price-Whelan, A.~M., Sip{\H{o}}cz, B.~M., et al.\ 2018, \aj, 156, 123.
\bibitem[Bailer-Jones et al.(2018)]{Bailer-Jones18} Bailer-Jones, C.~A.~L., Rybizki, J., Fouesneau, M., et al.\ 2018, \aj, 156, 58.
\bibitem[Balick \& Frank(2002)]{Balick02} Balick, B., \& Frank, A.\ 2002, \araa, 40, 439 
\bibitem[Balick(2004)]{Balick04} Balick, B.\ 2004, \aj, 127, 2262 
% \bibitem[Battinelli \& Demers(2006)]{Battinelli06} Battinelli, P., \& Demers, S.\ 2006, \aap, 447, 473
% \bibitem[Bertin \& Arnouts(1996)]{Bertin96} Bertin, E., \& Arnouts, S.\ 1996, \aaps, 117, 393 
\bibitem[Bil{\'\i}kov{\'a} et al.(2010)]{Bilikova10} Bil{\'\i}kov{\'a}, J., Chu, Y.-H., Gruendl, R.~A., et al.\ 2010, \aj, 140, 1433
\bibitem[Blackman, et al.(2001)]{Blackman01} Blackman, E.~G., Frank, A. \& Welch, C.\ 2001, \apj, 546, 288
% \bibitem[Boffin, et al.(2012)]{Boffin12} Boffin, H.~M.~J., Miszalski, B., Rauch, T., et al.\ 2012, Science, 338, 773
% \bibitem[Bond, et al.(2003)]{Bond03} Bond, H.~E., Pollacco, D.~L. \& Webbink, R.~F.\ 2003, \aj, 125, 260
% \bibitem[Bond, \& Ciardullo(2018)]{Bond18} Bond, H.~E., \& Ciardullo, R.\ 2018, Research Notes of the American Astronomical Society, 2, 143.
% \bibitem[B{\"o}hm-Vitense et al.(2000)]{Bohm-Vitense00} B{\"o}hm-Vitense, E., Carpenter, K., Robinson, R., Ake, T., \& Brown, J.\ 2000, \apj, 533, 969 
% \bibitem[Bothun et al.(1991)]{Bothun91}Bothun, G. et al. 1991, AJ, 101, 2220
\bibitem[Bromm, \& Loeb(2003)]{Bromm03} Bromm, V., \& Loeb, A.\ 2003, \nat, 425, 812
\bibitem[Bujarrabal et al.(2013)]{Bujarrabal13} Bujarrabal, V., Alcolea, J., Van Winckel, H., et al.\ 2013, \aap, 557, A104.
% %\bibitem[Catal{\'a}n et al.(2008)]{Catalan08} Catal{\'a}n, S., Isern, J., Garc{\'{\i}}a-Berro, E., et al.\ 2008, \aap, 477, 213 
\bibitem[Chabrier \& Baraffe(1997)]{Chabrier97} Chabrier, G. \& Baraffe, I.\ 1997, \aap, 327, 1039.
\bibitem[Chen et al.(2018)]{Chen18} Chen, Z., Blackman, E.~G., Nordhaus, J., Frank, A. \& Carroll-Nellenback, J. 2018, MNRAS, 473, 747
% %\bibitem[Christlieb et al.(2001)]{Christlieb01} Christlieb, N., Green, P.~J., Wisotzki, L., \& Reimers, D.\ 2001, \aap, 375, 366 
% %\bibitem[Clark et al.(2012)]{Clark12} Clark, B.~M., Blake, C.~H., \& Knapp, G.~R.\ 2012, \apj, 744, 119 
\bibitem[Corradi et al.(2011)]{Corradi11} Corradi, R.~L.~M., Sabin, L., Miszalski, B., et al.\ 2011, \mnras, 410, 1349
% \bibitem[Corrales(2015)]{Corrales15} Corrales, L.\ 2015, \apj, 805, 23 
% \bibitem[Covey et al.(2007)]{Covey07} Covey, K.~R., et al.\ 2007, \aj, 134, 2398 
% \bibitem[Covey et al.(2008)]{Covey08} Covey, K.~R., Ag{\"u}eros, M.~A., Green, P.~J., et al.\ 2008, \apjs, 178, 339 
\bibitem[Cutri et al.(2013)]{Cutr13} Cutri, R.~M. et al. \ 2013, VizieR Online Data Catalog, II/328
\bibitem[Dahn et al.(1977)]{Dahn77} Dahn, C.~C., Liebert, J., Kron, R.~G., Spinrad, H., \& Hintzen, P.~M.\ 1977, \apj, 216, 757 
% \bibitem[Dahn, et al.(2017)]{Dahn17} Dahn, C.~C., Harris, H.~C., Subasavage, J.~P., et al.\ 2017, \aj, 154, 147
% \bibitem[Davis (1987)]{Davis87} Davis, S. P. 1987, PASP, 99, 1105
\bibitem[Dearborn et al.(1986)]{Dearborn86} Dearborn, D.~S.~P., Liebert, J., Aaronson, M., et al.\ 1986, \apj, 300, 314 
% \bibitem[de Jong et al.(2010)]{deJong10} de Jong, J.~T.~A., Yanny, B., Rix, H.-W., et al.\ 2010, \apj, 714, 663 
\bibitem[de\,Kool \& Green(1995)]{deKool95} de Kool, M. \& Green, P.~J.\ 1995, \apj, 449, 236 
% \bibitem[De\,Marco et al.(2008)]{DeMarco08}De\,Marco O., Hillwig T. C., \& Smith A. J., 2008, AJ , 136, 323
% \bibitem[De Marco(2009)]{DeMarco09} De Marco, O.\ 2009, \pasp, 121, 316 
% \bibitem[Demers \& Battinelli(2007)]{Demers07} Demers, S., \& Battinelli, P.\ 2007, \aap, 473, 143 
\bibitem[Downes et al.(2004)]{Downes04} Downes, R.~A., Margon, B., Anderson, S.~F., et al.\ 2004, \aj, 127, 2838 
% %\bibitem[Duan et al.(2009)]{Duan09} Duan, F.-Q. et al, 2009, Research  in Astron. Astrophys, 9, 341 % Automated spectral classification using template matching
% \bibitem[Dufour et al.(2005)]{Dufour05}Dufour, P., Bergeron, P., \& Fontaine, G. 2005, ApJ, 627, 404
% \bibitem[Dufour et al.(2008)]{Dufour08}Dufour, P., Fontaine, G., Liebert, J., Schmidt, G.~D., \& Behara, N.\ 2008, \apj, 683, 978 
\bibitem[Evans et al.(2010)]{Evans10} Evans, I. N., et al. 2010, ApJS, 189, 37
% %\bibitem[Garc{\'e}s et al.(2011)]{Garces11} Garc{\'e}s, A., Catal{\'a}n, S., \& Ribas, I.\ 2011, \aap, 531, A7 
\bibitem[Farihi et al.(2018)]{Farihi18} Farihi, J., Arendt, A.~R., Machado, H.~S., \& Whitehouse, L.~J.\ 2018, \mnras, 477, 3801 
% \bibitem[Ferland, et al.(2013)]{Ferland13} Ferland, G.~J., Porter, R.~L., van Hoof, P.~A.~M., et al.\ 2013, \rmxaa, 49, 137
\bibitem[Filippazzo et al.(2015)]{Fili15} Filippazzo, J. C.; Rice, E. L.; Faherty, J.; Cruz, K. L.; Van Gordon, M. M.; Looper, D. L. \ 2015, \apj, 810, 158 
\bibitem[Fortney(2012)]{Fortney12} Fortney, J.~J.\ 2012, \apj, 747, L27.
\bibitem[Frew(2008)]{Frew08} Frew, D.~J.\ 2008, Ph.D. Thesis
\bibitem[Gaia Collaboration et al.(2018)]{Gaia18} Gaia Collaboration, Brown, A.~G.~A., Vallenari, A., et al.\ 2018, arXiv:1804.09365 
\bibitem[Garc{\'{\i}}a-Segura et al.(2018)]{Garcia18} Garc{\'{\i}}a-Segura, G., Ricker, P.~M., \& Taam, R.~E.\ 2018, \apj, 860, 19 
\bibitem[Garraffo et al.(2018)]{Garraffo18} Garraffo, C., Drake, J.~J., Alvarado-Gomez, J.~D., et al.\ 2018, \apj, 868, 60.
\bibitem[Gondoin(2018)]{Gondoin18} Gondoin, P.\ 2018, \aap, 616, A154 
% \bibitem[Gould \& Kollmeier(2004)]{Gould04} Gould, A., \& Kollmeier, J.~A.\ 2004, VizieR Online Data Catalog, 215, 20103 
% \bibitem[Gray \& Corbally(2009)]{Gray09} Gray, R.~O., \& Corbally, C.,  J.\ 2009, Stellar Spectral Classification by Richard O.~Gray and  Christopher J.~Corbally.~Princeton University Press, 2009. ISBN: 978-0-691-12511-4
% \bibitem[Gray et al.(2011)]{Gray11} Gray, R.~O., McGahee, C.~E., Griffin, R.~E.~M., \& Corbally, C.~J.\ 2011, \aj, 141, 160 
\bibitem[Green et al.(1991)]{Green91} Green, P.~J., Margon, B., \& MacConnell, D.~J.\ 1991, \apjl, 380, L31 %3 newly-recognized dCs
% %\bibitem[Green et al.(1992)]{Green92} Green, P.~J., Margon, B., Anderson, S.~F., \& MacConnell, D.~J.\ 1992, \apj, 400, 659 % Lum Ind
% \bibitem[Green \& Margon(1994)]{Green94} Green, P.~J., \& Margon, B.\ 1994, \apj, 423, 723 % Constraints on the Origin
\bibitem[Green(2013)]{Green13} Green, P.\ 2013, \apj, 765, 12 
\bibitem[Green(2018a)]{Green18a} Green, G. M. \ 2018, Journal of Open Source Software, 3(26), 695
% \bibitem[Green et al.(2018b)]{Green18b} Green, G.~M., Schlafly, E.~F., Finkbeiner, D., et al.\ 2018, \mnras, 478, 651 
% \bibitem[Hall \& Maxwell(2008)]{Hall08} Hall, P.~B., \& Maxwell, A.~J.\ 2008, \apj, 678, 1292 
% \bibitem[Harris et al.(1998)]{Harris98} Harris, H.~C., Dahn, C.~C., Walker, R.~L., et al.\ 1998, \apj, 502, 437 
% \bibitem[Harris et al.(2018)]{Harris18} Harris, H.~C., Dahn, C.~C., Subasavage, J.~P., et al.\ 2018, arXiv:1804.09824 
\bibitem[Heber et al.(1993)]{Heber93} Heber, U., Bade, N., Jordan, S., \&  Voges, W.\ 1993, \aap, 267, L31 % PG 0824+289 - A dwarf carbon star with a visible white dwarf companion
\bibitem[Iaconi et al.(2018)]{Iaconi18} Iaconi, R., De Marco, O., Passy, J.-C., et al.\ 2018, \mnras, 477, 2349.
% \bibitem[Izzard et al.(2009)]{Izzard09} Izzard, R.~G., Glebbeek, E., Stancliffe, R.~J., \& Pols, O.~R.\ 2009, \aap, 508, 1359 %	Population synthesis of binary CEMP stars
% \bibitem[Izzard et al.(2010)]{Izzard10} Izzard, R.~G., Dermine, T., \& Church, R.~P.\ 2010, \aap, 523, A10 
\bibitem[Jeffries \& Stevens(1996)]{Jeffries96} Jeffries, R.~D., \& Stevens, I.~R.\ 1996, \mnras, 279, 180 
\bibitem[Ji et al.(2016)]{Ji16} Ji, W., Cui, W., Liu, C., et al.\ 2016, \apjs, 226, 1 
% \bibitem[Jones D. \& Boffin H. M. J.(2017)]{Jones17}Jones D., Boffin H. M. J., 2017, Nature Astron. , 1, 0117
\bibitem[Jones et al.(2015)]{Jones15} Jones, D., Boffin, H.~M.~J., Rodr{\'{\i}}guez-Gil, P., et al.\ 2015, \aap, 580, A19 
\bibitem[Jones et al.(2017)]{Jones17} Jones, D., Van Winckel, H., Aller, A., Exter, K., \& De Marco, O.\ 2017, \aap, 600, L9 
% \bibitem[Joyce(1998)]{Joyce98} Joyce, R.~R.\ 1998, \aj, 115, 2059 
% \bibitem[Karachentsev et al.(2004)]{Karachentsev04}Karachentsev I.D., Karachentseva V.E., Huchtmeier W.K., \& Makarov D.I. 2004, AJ, 127, 2031
\bibitem[Kastner, et al.(2012)]{Kastner12} Kastner, J.~H., Montez, R., Balick, B., et al.\ 2012, \aj, 144, 58.
% \bibitem[Kinemuchi et al.(2008)]{Kinemuchi08} Kinemuchi, K., Harris, H.~C., Smith, H.~A., et al.\ 2008, \aj, 136, 1921
% \bibitem[Koester \& Knist(2006)]{Koester06} Koester, D., \& Knist, S.\ 2006, \aap, 454, 951 
\bibitem[Kosovichev et al.(2013)]{Kosovichev13} Kosovichev, A.~G., de Gouveia Dal Pino, E., \& Yan, Y.\ 2013, Solar and Astrophysical Dynamos and Magnetic Activity.
% \bibitem[Kowalski(2010)]{Kowalski10} Kowalski, P.~M.\ 2010, \aap, 519, L8 
\bibitem[Kraft(1967)]{Kraft67} Kraft, R.~P.\ 1967, \apj, 150, 551.
% \bibitem[Kraus \& Hillenbrand(2007)]{Kraus07} Kraus, A.~L., \& Hillenbrand , L.~A.\ 2007, \aj, 134, 2340 
% \bibitem[Laughlin, Bodenheimer \& Adams(1997)]{Laughlin97}Laughlin, G., Bodenheimer, P., \& Adams, F. C. 1997, ApJ, 482, 420
% \bibitem[Lee et al.(2008)]{Lee08} Lee, Y.~S., Beers, T.~C.,  Sivarani, T., et al.\ 2008, \aj, 136, 2022 
\bibitem[Li et al.(2018)]{Li18} Li, Y.-B., Luo, A.-L., Du, C.-D., et al.\ 2018, \apjs, 234, 31 
\bibitem[Liebert et al.(1994)]{Liebert94} Liebert, J., Schmidt, G.~D.,  Lesser, M., et al.\ 1994, \apj, 421, 733 %DA/dC CBS 311
% %\bibitem[Liu et al.(2012)]{Liu12} Liu, C., Li, L., Zhang, F., et al.\ 2012, \mnras, 424, 1841
% \bibitem[Lloyd Evans(2010)]{LloydEvans10} Lloyd Evans, T.\ 2010, Journal of Astrophysics and Astronomy, 31, 177 
% \bibitem[Lu(1991)]{Lu91} L\"u, P.K. 1991, AJ, 101, 2229
\bibitem[Lucatello et al.(2005)]{Lucatello05} Lucatello, S., Tsangarides, S., Beers, T.~C., Carretta, E., Gratton, R.~G., \& Ryan, S.~G.\ 2005, \apj, 625, 825 
\bibitem[Maldonado et al.(2017)]{Maldonado17} Maldonado, J., Scandariato, G., Stelzer, B., et al.\ 2017, \aap, 598, A27 
% \bibitem[Margon et al.(2002)]{Margon02} Margon, B., Anderson, S.~F., Harris, H.~C., et al.\ 2002, \aj, 124, 1651 
\bibitem[Margon et al.(2018)]{Margon18} Margon, B., Kupfer, T., Burdge, K., et al.\ 2018, \apjl, 856, L2 
\bibitem[Matt et al.(2015)]{Matt15} Matt, S.~P., Brun, A.~S., Baraffe, I., et al.\ 2015, \apj, 799, L23.
% \bibitem[Martin et al.(2005)]{Martin05} Martin D. C. et al., 2005, ApJ, 619, L1
\bibitem[Barrett et al.(2005)]{matplotlib05} Barrett, P., Hunter, J., Miller, J.~T., et al.\ 2005, Astronomical Data Analysis Software and Systems XIV, 91.
% \bibitem[McClure \& Woodsworth(1990)]{McClure90} McClure, R.~D., \& Woodsworth, A.~W.\ 1990, \apj, 352, 709 
% \bibitem[McClure(1997)]{McClure97} McClure, R.~D.\ 1997, \pasp,  109, 536 
% \bibitem[McConnachie(2012)]{McCon12} McConnachie, A.~W.\ 2012, \aj, 144, 4 
\bibitem[Micela et al.(1985)]{Micela85} Micela, G., Sciortino, S., Serio, S., et al.\ 1985, \apj, 292, 172 
% %\bibitem[Mizalski et al.(2012a)]{Mizalski12a} Mizalski, B., et al. 2012a, MNRAS, 419, 39
% %\bibitem[Mizalski et al.(2012b)]{Mizalski12b} Mizalski, B., Boffin, H.M.J, \& Corradi, R.L.M 2012, MNRAS, in press
\bibitem[Miszalski, et al.(2009)]{Miszalski09} Miszalski, B., Acker, A., Moffat, A.~F.~J., et al.\ 2009, \aap, 496, 813
\bibitem[Miszalski, et al.(2012)]{Miszalski12} Miszalski, B., Boffin, H.~M.~J., Frew, D.~J., et al.\ 2012, \mnras, 419, 39
\bibitem[Miszalski, et al.(2013)]{Miszalski13a} Miszalski, B., Boffin, H.~M.~J., \& Corradi, R.~L.~M.\ 2013, \mnras, 428, L39 
\bibitem[Miszalski et al.(2013)]{Miszalski13b} Miszalski, B., Boffin, H.~M.~J., Jones, D., et al.\ 2013, \mnras, 436, 3068.
\bibitem[Miszalski et al.(2018)]{Miszalski18} Miszalski, B., Manick, R., Miko{\l}ajewska, J., et al.\ 2018, \mnras, 473, 2275 
\bibitem[Miszalski et al.(2019)]{Miszalski19} Miszalski, B., Manick, R., Van Winckel, H., et al.\ 2019, \pasa, 36, e018.
\bibitem[Montez, et al.(2010)]{Montez10} Montez, R., De Marco, O., Kastner, J.~H., et al.\ 2010, \apj, 721, 1820.
\bibitem[Montez et al.(2015)]{Montez15} Montez, R., Jr., Kastner, J.~H., Balick, B., et al.\ 2015, \apj, 800, 8 
\bibitem[Morales et al.(2008)]{Morales08} Morales, J.~C., Ribas, I., \& Jordi, C.\ 2008, \aap, 478, 507 
\bibitem[Morgan et al.(2012)]{Morgan12} Morgan, D.~P., West, A.~A., Garc{\'e}s, A., et al.\ 2012, \aj, 144, 93
\bibitem[Morrison \& McCammon(1983)]{Morrison83} Morrison, R., \& McCammon, D.\ 1983, \apj, 270, 119 
% \bibitem[Mould et al.(1985)]{Mould85}Mould, J. et al. 1985, PASP, 97, 130
% \bibitem[Moultaka et al.(2004)]{Moultaka04} Moultaka, J., Ilovaisky, S.~A., Prugniel, P., \& Soubiran, C.\ 2004, \pasp, 116, 693 
% \bibitem[Mullan \& MacDonald(2001)]{Mullan01} Mullan, D.~J. \& MacDonald, J.\ 2001, \apj, 559, 353
% %\bibitem[Munari \& Renzini(1992)]{Munari92} Munari, U., \& Renzini, A.\ 1992, \apjl, 397, L87 
\bibitem[Munn et al.(2004)]{Munn04} Munn, J.~A., Monet, D.~G., Levine, S.~E., et al.\ 2004, \aj, 127, 3034 
\bibitem[Norris et al.(2013)]{Norris13} Norris, J.~E., Yong, D., Bessell, M.~S., et al.\ 2013, \apj, 762, 28
\bibitem[Noyes et al.(1984)]{Noyes84} Noyes, R.~W., Hartmann, L.~W., Baliunas, S.~L., Duncan, D.~K., \& Vaughan, A.~H.\ 1984, \apj, 279, 763 
\bibitem[Nutzman \& Charbonneau(2008)]{Nutzman08}Nutzman, P., \& Charbonneau, D. 2008, PASP, 120, 317
\bibitem[O'Dwyer, et al.(2003)]{O'Dwyer03} O'Dwyer, I.~J., Chu, Y.-H., Gruendl, R.~A., et al.\ 2003, \aj, 125, 2239
\bibitem[Oliphant(2006)]{NumPy06} Oliphant, T. E. 2006, \url{http://www.tramy.us/}
\bibitem[Oomen et al.(2018)]{Oomen18} Oomen, G.-M., Van Winckel, H., Pols, O., et al.\ 2018, \aap, 620, A85 
\bibitem[Pallavicini et al.(1981)]{Pallavicini81} Pallavicini, R., Golub, L., Rosner, R., et al.\ 1981, \apj, 248, 279
\bibitem[Reiners et al.(2012)]{Reiners12} Reiners, A., Joshi, N., \& Goldman, B.\ 2012, \aj, 143, 93.
\bibitem[Smith, R.K. et al.(2001)]{Smith01} Smith R. K., Brickhouse N. S., Liedahl D. A., Raymond J. C., 2001, ApJ , 556, L91  \url{https://heasarc.gsfc.nasa.gov/xanadu/xspec/manual/XSmodelApec.html}
% \bibitem[Pagano(2009)]{Pagano09} Pagano, I.\ 2009, \apss, 320, 115 
% %\bibitem[Parsons et al.(2012)]{Parsons12} Parsons, S.~G., G{\"a}nsicke, B.~T., Marsh, T.~R., et al.\ 2012, \mnras, 426, 1950 
% %\bibitem[Parsons et al.(2012)]{Parsons13} Parsons, S.~G., G{\"a}nsicke, B.~T., Marsh, T.~R., et al.\ 2012, arXiv:1211.0316 
\bibitem[Parker(1955)]{Parker55} Parker, E.~N.\ 1955, \apj, 122, 293 
% \bibitem[Plant et al.(2016)]{Plant16} Plant, K.~A., Margon, B., Guhathakurta, P., et al.\ 2016, \apj, 833, 232 
\bibitem[Plez \& Cohen(2005)]{Plez05} Plez, B., \& Cohen, J.~G.\ 2005, \aap, 434, 1117 
% %\bibitem[Pyrzas et al.(2011)]{Pyrzas11} Pyrzas, S., G{\"a}nsicke, B.~T., Brady, S., et al.\ 2011, \mnras, 1956 
% %\bibitem[Rebassa-Mansergas et al.(2010)]{Rebassa10} Rebassa-Mansergas, A., G{\"a}nsicke, B.~T., Schreiber, M.~R., Koester, D., \& Rodr{\'{\i}}guez-Gil, P.\ 2010, \mnras, 402, 620 
%\bibitem[Reiners \& Basri(2008)]{Reiners08}Reiners, A., \& Basri, G 2008, ApJ, 684, 1390
% \bibitem[Rebassa-Mansergas et al.(2012)]{Rebassa12} Rebassa-Mansergas, A., Nebot G{\'o}mez-Mor{\'a}n, A., Schreiber, M.~R., et al.\ 2012, \mnras, 419, 806 
% \bibitem[Rebassa-Mansergas et al.(2016)]{Rebassa16} Rebassa-Mansergas, A., Ren, J.~J., Parsons, S.~G., et al.\ 2016, \mnras, 458, 3808 
\bibitem[Reichardt et al.(2019)]{Reichardt19} Reichardt, T.~A., De Marco, O., Iaconi, R., et al.\ 2019, \mnras, 484, 631.
% \bibitem[Reggiani \& Meyer(2011)]{Reggiani11} Reggiani, M.~M., \& Meyer, M.~R.\ 2011, \apj, 738, 60 
\bibitem[Ricker et al.(2015)]{Ricker15} Ricker, G.~R., Winn, J.~N., Vanderspek, R., et al.\ 2015, Journal of Astronomical Telescopes, Instruments, and Systems, 1, 14003.
\bibitem[Rosen et al.(2016)]{Rosen16} Rosen, S.~R., Webb, N.~A., Watson, M.~G., et al.\ 2016, \aap, 590, A1 
\bibitem[Roulston et al.(2019)]{Roulston19} Roulston, B.~R., Green, P.~J., Ruan, J.~J., et al.\ 2019, \apj, 877, 44.
% \bibitem[Salim \& Gould(2003)]{Salim03} Salim, S. \& Gould, A. 2003,  ApJ, 582, 1000
% \bibitem[Schwarz et al.(2011)]{Schwarz11} Schwarz, G.~J., Ness, J.-U., Osborne, J.~P., et al.\ 2011, \apjs, 197, 31
% \bibitem[Jones, E., Oliphant, T, Peterson, P. et al.(2001)]{scipy} Jones, E., Oliphant, T, Peterson, P. et al.\ 2001--, \url{http://www.scipy.org/}
% %\bibitem[Silvestri et al.(2006)]{Silvestri06}Silvestri, N. M., et al., 2006, AJ, 131, 1674
% \bibitem[Secchi(1869)]{Secchi69} Secchi, A.\ 1869, Astronomische Nachrichten, 73, 129
\bibitem[Si et al.(2014)]{Si14} Si, J., Luo, A., Li, Y., et al.\ 2014, Science China Physics, Mechanics, and Astronomy, 57, 176 
\bibitem[Skrutskie et al.(2006)]{Skru06} Skrutskie, M. F.; Cutri, R. M.; Stiening, R.; Weinberg, M. D.; Schneider, S.; Carpenter, J. M.; Beichman, C.; Capps, R.; Chester, T.; Elias, J.; Huchra, J.; Liebert, J.; Lonsdale, C.; Monet, D. G.; Price, S.; Seitzer, P.; Jarrett, T.; Kirkpatrick, J. D.; Gizis, J. E.; Howard, E.; Evans, T.; Fowler, J.; Fullmer, L.; Hurt, R., Light, R.; Kopan, E. L.; Marsh, K. A.; McCallon, H. L.; Tam, R.; Van Dyk, S.; Wheelock, S.\ 2006, \aj, 131, 1163 
\bibitem[Skumanich(1972)]{Skumanich72} Skumanich, A.\ 1972, \apj, 171, 565 
\bibitem[Smith et al.(2007)]{Smith07} Smith, M.~C., et al.\ 2007, \mnras, 379, 755 
\bibitem[Soker \& Kastner(2002)]{Soker02} Soker, N., \& Kastner, J.~H.\ 2002, \apj, 570, 245
% \bibitem[Stanghellini, et al.(2017)]{Stanghellini17} Stanghellini, L., Bucciarelli, B., Lattanzi, M.~G., et al.\ 2017, \na, 57, 6
\bibitem[Starkenburg et al.(2014)]{Starkenburg14} Starkenburg, E., Shetrone, M.~D., McConnachie, A.~W., et al.\ 2014, \mnras, 441, 1217
\bibitem[Stelzer et al.(2013)]{Stelzer13} Stelzer, B., Marino, A., Micela, G., et al.\ 2013, \mnras, 431, 2063 
\bibitem[Stelzer et al.(2016)]{Stelzer16} Stelzer, B., Damasso, M., Scholz, A., Matt, S.P.  2016, \mnras, 431, 2063 
% \bibitem[Stoughton et al.(2002)]{Stoughton02} Stoughton, C., Lupton, R.~H., Bernardi, M., et al.\ 2002, \aj, 123, 485 
% \bibitem[Str\"omberg(1939)]{Stromberg39} Str\"omberg, G. 1939, ApJ, 89, 10
% %\bibitem[Tappert et al.(2011a)]{Tappert11a} Tappert, C., G{\"a}nsicke, Rebassa-Mansergas, A., Schmidtobreick, L., \& Schreiber, M. R. 2011, A\&A, 531, A113
% %\bibitem[Tappert et al.(2011b)]{Tappert11b} Tappert, C., G{\"a}nsicke, B.~T.,  Schmidtobreick, L., \& Ribeiro, T.\ 2011, \aap, 532, A129 
\bibitem[Tielens(2005)]{Tielens05} Tielens, A.~G.~G.~M.\ 2005, The Physics and Chemistry of the Interstellar Medium.
\bibitem[Testa et al.(2004)]{Testa04} Testa, P., Drake, J.~J., \& Peres, G.\ 2004, \apj, 617, 508.
% \bibitem[Tonry \& Davis(1979)]{Tonry79} Tonry, J., \& Davis, M.\ 1979, \aj, 84, 1511 
\bibitem[Tyndall et al.(2014)]{Tyndall14} Tyndall, A.~A., Jones, D., Boffin, H.~M.~J., et al.\ 2014, Revista Mexicana De Astronomia Y Astrofisica Conference Series, 60.
% \bibitem[van den Bergh(2000)]{vdB00} van den Bergh, S.\ 2000, \pasp, 112, 529 
% \bibitem[Vanture (1992)]{Vanture92} Vanture, A. D. 1992, AJ, 104, 1997
% \bibitem[van Winckel et al.(2009)]{VanWinckel09} van Winckel, H., Lloyd Evans, T., Briquet, M., et al.\ 2009, \aap, 505, 1221 
% \bibitem[Van Winckel et al.(2014)]{VanWinckel14} Van Winckel, H., Jorissen, A., Exter, K., et al.\ 2014, \aap, 563, L10 
% \bibitem[Voges et al.(2000)]{Voges00} Voges, W., et al. 2000, VizieR Online Data Catalog, 9029, 0
% \bibitem[Vogt et al.(1995)]{Vogt95} Vogt, S.~S., Mateo, M., Olszewski, E.~W., \& Keane, M.~J.\ 1995, \aj, 109, 151 
% \bibitem[Vogt, et al.(2014)]{Vogt14} Vogt, F.~P.~A., Dopita, M.~A., Kewley, L.~J., et al.\ 2014, \apj, 793, 127
% \bibitem[Waelkens et al.(1991)]{Waelkens91}Waelkens C., VanWinkel H., Boegaert E., Trams  N.R. 1991 \aap, 251, 495
% \bibitem[Wallerstein \& Knapp(1998)]{Wall98} Wallerstein, G., \& Knapp, G.~R.\ 1998, \araa, 36, 369 
% \bibitem[Watson et al.(2009)]{Watson09} Watson, M.~G., Schr{\"o}der, A.~C., Fyfe, D., et al.\ 2009, \aap, 493, 339 
\bibitem[West et al.(2008)]{West08} West, A.~A., Hawley, S.~L., Bochanski, J.~J., et al.\ 2008, \aj, 135, 785 
\bibitem[Whitehouse et al.(2018)]{Whitehouse18}Whitehouse, L.~J., Farihi, J., Green, P. J., Wilson, T.G, \& Subasavage, J.P. 2018, \mnras, 479, 3873
% %\bibitem[Willems \& Kolb(2004)]{Willems04} Willems, B., \& Kolb, U.\ 2004, \aap, 419, 1057 
% \bibitem[Wilson \& Nordhaus(2019)]{Wilson19} Wilson, E.~C. \& Nordhaus, J.\ 2019, \mnras, 485, 4492.
\bibitem[Wood et al.(2004)]{Wood04} Wood, P.~R., Olivier, E.~A., \& Kawaler, S.~D.\ 2004, \apj, 604, 800 
% \bibitem[Wright et al.(2010)]{Wright10} Wright, E.~L.,  Eisenhardt, P.~R.~M., Mainzer, A.~K., et al.\ 2010, \aj, 140, 1868 
\bibitem[Wright et al.(2010)]{Wright10} Wright, N.~J., Drake, J.~J., \& Civano, F.\ 2010, \apj, 725, 480.
\bibitem[Wright et al.(2011)]{Wright11} Wright, N.~J., Drake, J.~J., Mamajek, E.~E., \& Henry, G.~W.\ 2011, \apj, 743, 48 
\bibitem[Wright et al.(2018)]{Wright18} Wright, N.~J., Newton, E.~R., Williams, P.~K.~G., Drake, J.~J., \& Yadav, R.~K.\ 2018, \mnras, 479, 2351 
\bibitem[Yoon et al.(2016)]{Yoon16} Yoon, J., Beers, T.~C., Placco, V.~M., et al.\ 2016, \apj, 833, 20 
% \bibitem[York et al.(2000)]{York00} York, D.~G., et al.\ 2000, \aj, 120, 1579
% \bibitem[Zamora et al.(2009)]{Zamora09} Zamora, O., Abia, C., Plez, B., Dom{\'{\i}}nguez, I., \& Cristallo, S.\ 2009, \aap, 508, 909 
% \bibitem[Zucker et al.(2006)]{Zucker06}Zucker, D.~B., Belokurov, V., Evans, N.~W., et al.\ 2006, \apjl, 643, L103
\end{thebibliography}



\end{document}